%% file: main.tex
\documentclass[conference,compsoc]{IEEEtran}
\usepackage{xurl}

\ifCLASSOPTIONcompsoc
  \usepackage[nocompress]{cite}
\else
  \usepackage{cite}
\fi

\input{macro}

\begin{document}

\title{When Being Wrong Is Right:\\
Reverse-Capability Audio CAPTCHAs through Psychoacoustic Design}

\author{
\IEEEauthorblockN{
Mingjie Bao\IEEEauthorrefmark{1},
Yu Dai\IEEEauthorrefmark{1},
Guangke Chen\IEEEauthorrefmark{1}\IEEEauthorrefmark{3},
Guozhu Meng\IEEEauthorrefmark{2}
}
\IEEEauthorblockA{
\IEEEauthorrefmark{1}Wuhan University 
\IEEEauthorrefmark{2}Institute of Information Engineering,
Chinese Academy of Sciences 
\IEEEauthorrefmark{3}Corresponding author
}
}
\maketitle

\begin{abstract}
Large audio-language models (LALMs) have recently been shown to bypass audio CAPTCHAs designed around limitations of automatic speech recognition.
To address this threat,
we introduce \emph{reverse-capability verification},
a general CAPTCHA framework that reverses the conventional human-easy, machine-hard relationship with respect to the physical ground truth. It selects challenges in which stable human perceptual limitations cause the human-reference answer to differ from the physical ground truth, while machine analysis tends toward that truth. Because the verifier accepts the human-reference answer, advanced solver capabilities and resulting improved recovery of the physical ground truth
may instead strengthen the human--machine separation.

We instantiate the framework through psychoacoustics and design two audio CAPTCHAs: \textsc{Hipi}, a classification task, and \textsc{Yingchen}, a counting task, together with their conjunctive composition. Across seven
LALMs and eight attack strategies, the maximum observed machine acceptance rates are 19\%, 17\%, and 3\% for \textsc{Hipi}, \textsc{Yingchen}, and their composition, respectively, compared with 45--100\% for nine prior schemes. A multidimensional human usability study yields corresponding human acceptance rates of 78\%, 90\%, and 76\%, while maintaining competitive efficiency and satisfaction. Further analysis demonstrates that advanced LALMs tend toward physical ground truth. These results show that reverse-capability verification substantially reduces machine acceptance while retaining high human acceptance through an approach different from conventional CAPTCHAs.
\end{abstract}

\input{introduction_new}
\input{02_background}

\input{03_problem_formulation}

\input{04_empirical_study}

\input{05_methodology}
\input{06_evaluation}

\input{discussion_new}

\input{conclusion_new}

\bibliographystyle{IEEEtran}
\bibliography{references}

\appendices

\input{appendices}

\clearpage

\end{document}

%% file: macro.tex
\usepackage{amsmath,amssymb}
\usepackage{graphicx}
\usepackage{tikz}
\usetikzlibrary{arrows.meta,positioning,calc,fit}
\usepackage{ragged2e}

\usepackage{array}
\usepackage{booktabs}
\usepackage{multirow}
\usepackage{tabularx}
\usepackage{threeparttable}
\usepackage{diagbox}
\usepackage{makecell}
\usepackage{float}

\usepackage{xcolor}
\usepackage{pifont}
\usepackage{xspace}
\usepackage{enumitem}
\usepackage{microtype}
\usepackage{url}

\usepackage[most]{tcolorbox}

\usepackage[hidelinks]{hyperref}
\usepackage{cleveref}
\crefname{section}{§}{§§}
\Crefname{section}{§}{§§}

\newcommand{\cmark}{\ding{51}}
\newcommand{\xmark}{\ding{55}}

\definecolor{HumanBlue}{RGB}{37,99,161}
\definecolor{MachineOrange}{RGB}{204,102,0}
\definecolor{PhysicalGray}{RGB}{90,90,90}
\definecolor{SoftGray}{RGB}{242,242,242}

\newcommand{\MARmy}[3]{%
  \makebox[1.25em][c]{#1}%
  \hspace{1pt}/\hspace{1pt}%
  \makebox[1.25em][c]{#2}%
  \hspace{1pt}/\hspace{1pt}%
  \makebox[1.25em][c]{#3}%
}

\newcommand{\ao}{\textsc{Hipi}\xspace}
\newcommand{\at}{\textsc{Yingchen}\xspace}

\newtcolorbox{promptbox}{
  breakable,
  colback=gray!10,
  colframe=gray!10,
  boxrule=0pt,
  arc=0pt,
  outer arc=0pt,
  left=4pt,
  right=4pt,
  top=3pt,
  bottom=3pt,
  boxsep=0pt,
  before skip=4pt,
  after skip=6pt,
  fontupper=\small
}

%% file: introduction_new.tex
\section{Introduction}
\label{sec:introduction}

Completely Automated Public Turing tests to tell Computers and Humans Apart (CAPTCHAs) are automatically generated challenge--response tests that protect online services by distinguishing legitimate users from automated programs~\cite{vonAhn2003captcha}. They gate operations such as account registration and content submission to limit automated abuse, including spam, scraping, and fraudulent activity. Most explicit CAPTCHA challenges, however, rely on visual perception, creating an accessibility barrier for people with visual impairments\cite{fanelle2020accessibility,guerar2021captcha}. Audio CAPTCHAs provide an alternative modality through which users listen and respond rather than interpret an image~\cite{fanelle2020accessibility}. They remain supported in major commercial services, including Google reCAPTCHA and {AWS WAF CAPTCHA}~\cite{googleRecaptchaAccessibility,awsWafCaptcha}.

The security of audio CAPTCHAs has developed through a continuing attack--defense arms race. Early designs asked users to transcribe spoken digits, letters, or words and added handcrafted interference, such as background noise, overlapping speech, filtering, or changes in playback speed, to impede automatic speech recognition (ASR)~\cite{kochanski2002reverse,bursztein2011failure}. As ASR improved, automated attacks were shown to break many such schemes, including widely deployed commercial systems~\cite{bursztein2011failure}. Designers subsequently pursued two main alternatives. Adversarial transcription CAPTCHAs replace handcrafted interference with perturbations deliberately optimized to disrupt ASR output~\cite{hossen2022aaecaptcha,abdullah2023yeehaw,wang2024adversarial,lee2024advcaptcha}. Reasoning-based CAPTCHAs instead require an operation beyond literal transcription, such as counting, classification, arithmetic, or selective reporting~\cite{fanelle2020accessibility,meutzner2017improved}. Because the former explicitly target recognition models and the latter require capabilities beyond transcription alone, both showed strong resistance under the ASR-centered threat models for which they were designed.

Large audio-language models (LALMs) fundamentally change this threat model. By coupling general audio perception with language understanding, instruction following, and reasoning, LALMs can process speech and non-speech audio together with a task instruction and directly produce the required answer~\cite{openai2026gptrealtime2,qwen2026qwen35omni}. A recent evaluation shows that LALM-based solvers solve reasoning-based audio CAPTCHAs beyond the scope of ASR and bypass schemes previously shown to resist ASR attacks~\cite{ding2026robustcaptchausingaudio}.
These findings establish the need for audio CAPTCHA designs that can withstand this stronger class of solvers. 
Such designs have one underlying challenge: audio CAPTCHAs typically follow the conventional premise of finding tasks that are easy for humans but difficult for machines\cite{vonAhn2003captcha}. As machines become increasingly capable of perceiving and reasoning about the physical world, sustaining this asymmetry becomes progressively harder. This raises a more fundamental question: 
\emph{must CAPTCHA security rely on machines being worse than humans at recovering the physical ground truth?}

We answer this question in the negative by introducing the \emph{reverse-capability verification} framework for CAPTCHA design. Conventional CAPTCHAs use the physical ground truth $y^\star$ as the verifier-accepted answer and therefore depend on humans recovering it more reliably than machines. Our framework instead selects challenges for which a stable human perceptual limitation causes the modal human response, or human-reference answer $y_H^\dagger$, to differ from $y^\star$, while machine analysis tends toward the physical ground truth. The verifier accepts $y_H^\dagger$ rather than $y^\star$, so physically accurate machine analysis produces a rejected response. 
This new class of CAPTCHA designs addresses the above challenge: advances in machine capabilities, and the resulting improvements in recovering the physical ground truth, need not weaken CAPTCHA security and may instead strengthen the human--machine separation. 

We instantiate this framework through psychoacoustics, which characterizes divergences between objectively measurable sound properties and human auditory perception~\cite{bregman1990auditoryscene,moore2012psychologyhearing}. The main challenge is that psychoacoustics encompasses many phenomena, not all of which elicit responses that are sufficiently stable 
and composable for CAPTCHA construction. We therefore follow the two-stage process summarized in \figurename~\ref{fig:framework-overview}. 
First, we systematically organize 13 candidate stimulus classes derived from nine base stimulus designs into perceptual illusion, perceptual priority, and perceptual boundary, and use a 
pilot study to assess human-response stability, finally retaining nine stimulus classes.  
Second, based on the retained stimulus classes, we design two concrete audio CAPTCHAs that integrate different compatible sets of classes. \textsc{Hipi} uses a classification task, whereas \textsc{Yingchen} uses a counting task. 
The two CAPTCHAs can also be composed conjunctively for stronger verification.

We evaluate the proposed CAPTCHAs through a comprehensive security evaluation and a multidimensional human usability study. 
For security against LALMs, 
we evaluate seven general-purpose LALMs, eight attack strategies ranging from direct inference to CAPTCHA-adaptive prompting and response aggregation, nine prior audio CAPTCHA baselines, and 100 independently generated challenges per proposed task. Across all evaluated solver--strategy configurations, the maximum machine acceptance rates (MARs) are 19\% for \textsc{Hipi}, 17\% for \textsc{Yingchen}, and 3\% for their composition, whereas the strongest attacks against the baselines reach 45--100\%. Our human study evaluates effectiveness, efficiency, and satisfaction through human acceptance rate (HAR), completion time, replay rate, and satisfaction ratings. 
\textsc{Hipi} and \textsc{Yingchen} achieve HARs of 78\% and 90\%, respectively. Notably, \textsc{Yingchen} achieves the lowest replay rate (12\%) among all evaluated CAPTCHAs, while both tasks maintain competitive satisfaction. 

\noindent\textbf{Contributions.}
We make the following contributions:
\begin{itemize}[leftmargin=*]
\item We introduce and formalize \emph{reverse-capability verification}, a general CAPTCHA design framework that decouples the verifier-accepted human response from the physical ground truth. It shows how challenges whose physical properties are easier for machines than for humans can still reliably distinguish the two.

\item We develop a systematic psychoacoustic instantiation of the framework. We organize and screen 13 candidate stimulus classes under three design principles, retain nine through a pilot study and integratability analysis, and use them to construct the classification-based \textsc{Hipi}, the counting-based \textsc{Yingchen}, and their conjunctive composition.

\item We conduct a broad security and usability evaluation comprising seven LALMs, eight inference-time attack strategies, nine baseline CAPTCHA schemes, and 
multidimensional human studies spanning effectiveness, efficiency, and satisfaction. 
The results demonstrate substantially lower machine acceptance than the baselines while maintaining high human acceptance. 

\end{itemize}

%% file: 02_background.tex
\section{Background}
\label{sec:background}

\subsection{Audio CAPTCHAs}
\label{sec:bg_captcha}

CAPTCHAs are automatically generated challenge--response tests designed to distinguish legitimate human users from automated solvers~\cite{vonAhn2003captcha}. 
We focus on explicit challenges whose submitted responses are accepted or rejected by a verifier, rather than a risk-scoring system. 
Operationally, a useful CAPTCHA must admit legitimate users at a high rate while rejecting automated solvers~\cite{vonAhn2003captcha,bursztein2011failure}.

Audio CAPTCHAs present challenges through audio and have long served as an alternative to visual CAPTCHAs for people with visual impairments (PVIs)~\cite{fanelle2020accessibility}. 
Existing audio CAPTCHAs broadly follow two task structures~\cite{ding2026robustcaptchausingaudio}. 
\emph{Transcription-based CAPTCHAs} require users to transcribe spoken digits, letters, words, or phrases presented with handcrafted distortions or adversarial perturbations intended to impede automated solvers~\cite{abdullah2023yeehaw,wang2024adversarial,lee2024advcaptcha}. \emph{Reasoning-based CAPTCHAs} require an operation beyond literal transcription, such as counting, classification, arithmetic, instruction following, or selective reporting~\cite{fanelle2020accessibility,meutzner2017improved,ding2026robustcaptchausingaudio}. 

Within these two broad task structures, related work has explored several distinct design strategies, including noisy speech transcription, adversarially perturbed audio, rule-based tasks, and perception-mismatch designs. 

\noindent\textbf{Noisy Speech Transcription-based CAPTCHAs.}
Traditional audio CAPTCHAs ask users to transcribe spoken digits, letters, or words while adding noise or signal distortion to reduce automatic recognition accuracy\cite{kochanski2002reverse,bursztein2011failure}. Early work examined transformations such as additive noise, echo, and filtering~\cite{kochanski2002reverse}, and subsequent systems extended this approach with overlapping speech, reverberation, changes in playback speed, and other manually selected distortions\cite{meutzner2014asr,fanelle2020accessibility}. 
Fanelle et al.'s \emph{Control} design emulates an alphanumeric CAPTCHA with variations in pitch, speed, amplification, and background noise. Their \emph{Pauses} design retains the same transcription task but inserts a two-second interval between
characters~\cite{fanelle2020accessibility}.

\noindent\textbf{Adversarial Audio-based CAPTCHAs.}
Later work replaces heuristically selected interference with perturbations explicitly designed to exploit ASR weaknesses while preserving human intelligibility~\cite{hossen2022aaecaptcha,abdullah2023yeehaw,wang2024adversarial,lee2024advcaptcha}. The design \emph{aaeCAPTCHA} applies projected gradient descent directly to the speech waveform, maximizing the CTC transcription loss of a target model under a bounded perturbation~\cite{hossen2022aaecaptcha}. \emph{Yeehaw Junction} builds on frequency-domain decimation by removing low-intensity spectral bins, clipping large amplitudes at dominant frequencies while preserving their peak locations, and injecting Gaussian noise~\cite{abdullah2023yeehaw}. 

\noindent\textbf{Rule-based CAPTCHAs.}
Rule-based designs require users to perceive the audio content and then apply an explicit operation, such as calculation, counting, classification, or event selection, to derive the submitted answer~\cite{fanelle2020accessibility,meutzner2016nonspeech,huang2021psychoacoustic,lazar2012soundsright}. Fanelle et al.~\cite{fanelle2020accessibility} provide several such designs. In \emph{Math}, users maintain a running total over a sequence of single-digit additions and subtractions. In \emph{Character}, they count the occurrences of a specified character in an alphanumeric sequence. In \emph{Categories}, they count sounds belonging to a target semantic category, such as birds, within a sequence containing unrelated environmental sounds.

\noindent\textbf{Perception-Mismatch-based CAPTCHAs.}
A further line of work deliberately constructs audio for which humans and automated systems recover different effective content~\cite{meutzner2015differences,choi2018poster,lee2024advcaptcha}. Meutzner et al.~\cite{meutzner2015differences,meutzner2017improved} present sequences containing meaningful words and acoustically similar speech-like nonsense, asking users to report only the meaningful words. Human listeners can use language understanding to distinguish the two, whereas ASR systems may transcribe both. Choi et al.~\cite{choi2018poster} adopt a hybrid CAPTCHA that mixes human-only and machine-only speech within the same challenge. \emph{Volcano}~\cite{lee2024advcaptcha} creates a related mismatch by concatenating a human-only digit and a machine-only digit. The human-only digit is generated through short-time Fourier transform (STFT)-based component removal until the target ASR no longer transcribes it, while the machine-only digit is produced by reducing the volume of a clean digit to a level that remains ASR-transcribable but is barely comprehensible to humans. Human listeners and ASR systems therefore recover different digit sequences. 
{Concurrent work by Ding et al.~\cite{ding2026robustcaptchausingaudio}
proposes IllusionAudio, which exploits sine-wave speech
illusions to preserve human speech intelligibility while
impairing automated recognition.}

Despite their different mechanisms, the designs above retain the conventional human-easy, machine-hard component: an automated solver must fail to transcribe, reason over, or select the intended content. This asymmetry is increasingly difficult to sustain as modern LALMs extend automated solving beyond transcription to general audio perception and reasoning~\cite{ding2026robustcaptchausingaudio}.
In contrast, our reverse-capability framework selects challenges where machine analysis tends toward the physical ground truth, while ordinary listeners reliably produce a different human-reference answer accepted by the verifier. We formalize this distinction in Section~\ref{sec:formulation-threat-goals}.

\subsection{Modern Audio CAPTCHA Solvers}
\label{sec:bg_solvers}

Modern automated solvers for audio CAPTCHAs extend beyond conventional automatic speech recognition (ASR). 

\emph{Large audio-language models} (LALMs) are multimodal large language models that process audio alongside natural-language instructions and generate textual or spoken responses~\cite{yang2025lalmsurvey}. Within a unified framework, LALMs combine audio perception, language understanding, instruction following, and reasoning, enabling them to recognize speech, interpret non-speech sounds, and reason over acoustic information~\cite{su2025audiolanguage}. These capabilities allow LALMs to process an audio CAPTCHA together with its task instructions and directly generate the required answer, making them general-purpose audio CAPTCHA solvers rather than transcription-only systems~\cite{ding2026robustcaptchausingaudio}. 
Recent work shows that LALM-based solvers can solve
non-transcription CAPTCHA types beyond the scope of ASR
and bypass audio CAPTCHAs previously shown to resist
ASR-based attacks~\cite{ding2026robustcaptchausingaudio}.
{Representative models include
GPT-Realtime-2~\cite{openai2026gptrealtime2},
Qwen3.5-Omni~\cite{qwen2026qwen35omni}, and Kimi-Audio~\cite{kimi2025audio}.}

\subsection{Psychoacoustics}
\label{sec:bg_psychoacoustics}

Psychoacoustics characterizes the relationship between the objective acoustic properties of sound and the subjective auditory sensations they evoke~\cite{moore2012psychologyhearing}. Importantly, hearing is not a direct readout of the acoustic signal: the auditory system filters and organizes incoming sounds into perceptual streams~\cite{bregman1990auditoryscene}. Consequently, a listener's percept may diverge from, or even contradict, the physical properties of the signal.

In this work, we develop a psychoacoustics-based approach to designing reverse-capability audio CAPTCHAs. Drawing on the psychoacoustic literature, we identify candidate stimulus classes and organize them into three categories: \emph{perceptual illusion}, \emph{perceptual priority}, and \emph{perceptual boundary}. We then evaluate multiple stimulus classes within each category as potential CAPTCHA building blocks and examine how compatible classes can be combined to construct effective reverse-capability audio CAPTCHAs.

%% file: 03_problem_formulation.tex
\section{Reverse-Capability Framework}
\label{sec:formulation-threat-goals}

We present our reverse-capability framework for 
CAPTCHA design, first formalizing the general setting and then contrasting conventional and reverse-capability CAPTCHAs. 

\noindent\textbf{Audio CAPTCHA formulation.}
Each CAPTCHA challenge $x\sim\mathcal{D}$ can be viewed as a question associated with a \emph{physical ground truth} $y^\star(x)\in\mathcal{Y}$, defined as the answer reflecting the objectively measurable property or attribute of the challenge stimulus. For an audio challenge, it may be the transcription of spoken content, the number of physically present sound events, or the presence of a target acoustic component. 

Let $q_H(y\mid x)$ and $q_{S,a}(y\mid x)$ denote the response distributions of humans and of solver $S$ using strategy $a$, respectively. 
Accordingly, $y_H(x)$ denotes a human response drawn from $q_H(\cdot\mid x)$, while $y_{S,a}(x)$ denotes a solver response drawn from $q_{S,a}(\cdot\mid x)$. 
We use $y_H^\dagger(x)$ to denote the modal human (i.e., most likely) response under $q_H(\cdot\mid x)$ and term it the \emph{human-reference answer}.

The verifier uses $y_H^\dagger(x)$
as its reference answer. For a parsed response $y\in\mathcal{Y}$, the acceptance function is
\begin{equation}
A(x,y)
=
\mathbb{I}\left[
y=y_H^\dagger(x)
\right]
\in\{0,1\}.
\label{eq:verifier}
\end{equation}

\begin{figure}[t]
    \centering
    \includegraphics[width=\columnwidth]{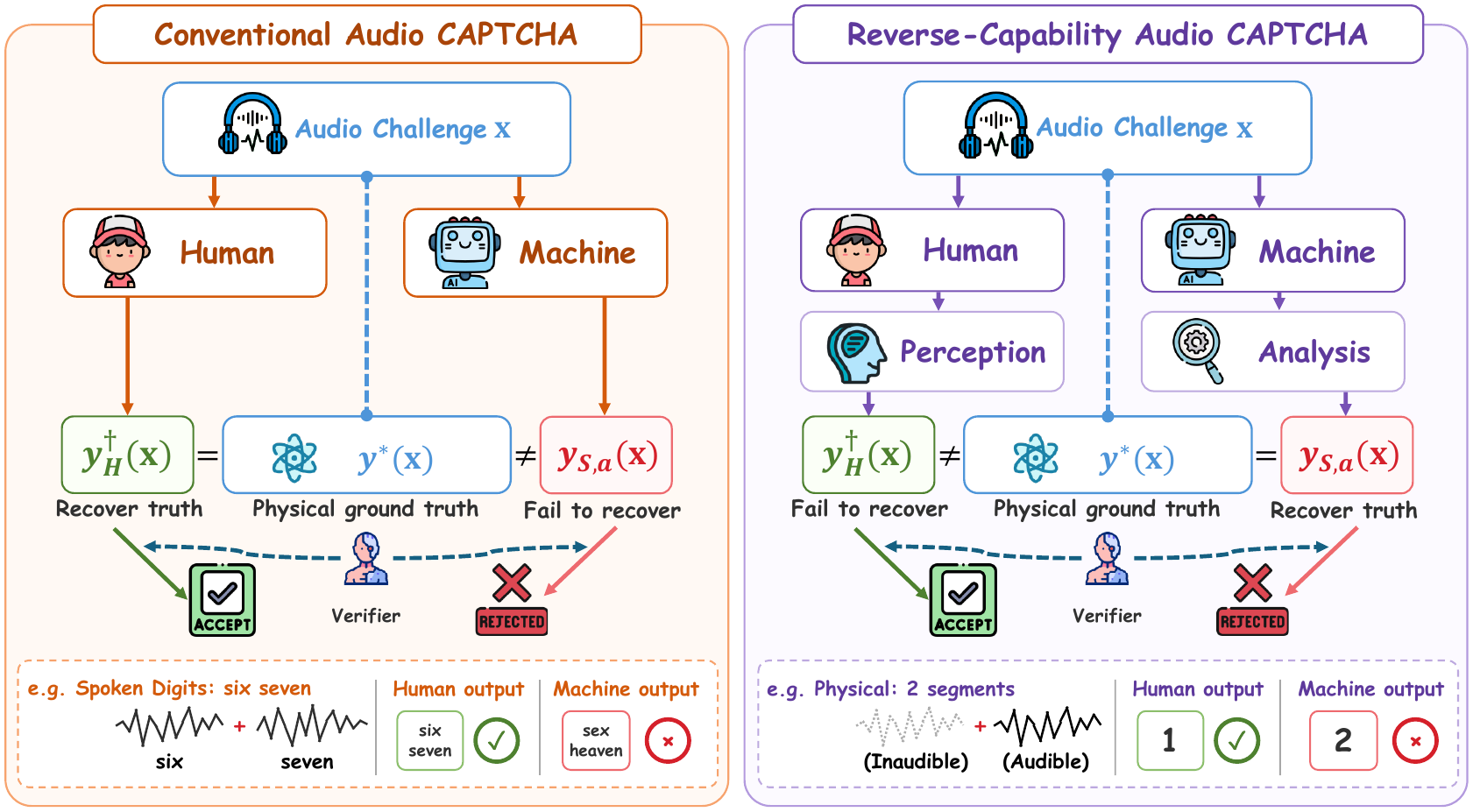}
    \caption{Comparison between conventional and reverse-capability CAPTCHAs.
    }
    \label{fig:reverse-capability-comparison}
\end{figure}

Thus, an ideal CAPTCHA that reliably accepts human users while rejecting automated solvers should satisfy
\begin{equation}
q_H\left(y_H^\dagger(x)\mid x\right)\approx 1,
\qquad
q_{S,a}\left(y_H^\dagger(x)\mid x\right)\approx 0.
\label{eq:human-machine-gap}
\end{equation}
Equation~\eqref{eq:human-machine-gap} captures the separation requirement shared by conventional and reverse-capability audio CAPTCHAs: the verifier-accepted response should be highly probable for humans but unlikely for automated solvers. The two paradigms differ in how the human and solver responses relate to the physical ground truth $y^\star(x)$, as demonstrated in \figurename~\ref{fig:reverse-capability-comparison}. 

\noindent\textbf{Conventional Audio CAPTCHAs}
follow the principle that a challenge should be easy for humans but difficult for machines~\cite{vonAhn2003captcha}. They therefore employ tasks whose physical ground truth can be recovered reliably by humans but not by automated solvers. Ideally,
\begin{equation}
q_H\bigl(y^\star(x)\mid x\bigr)\approx 1,
\qquad
q_{S,a}\bigl(y^\star(x)\mid x\bigr)\approx 0.
\label{eq:traditional-captcha}
\end{equation}
Together with Equation~\eqref{eq:human-machine-gap}, 
the human-reference answer coincides with the physical ground truth, i.e., $y_H^\dagger(x)=y^\star(x)$. 
The verifier can therefore use $y^\star(x)$ directly as its 
reference.

\noindent\textbf{Reverse-capability audio CAPTCHAs}
invert the above relationship to the physical ground truth. Specifically, they exploit a stable human perceptual limitation that causes human responses to systematically diverge from the physical ground truth, while direct machine analysis tends to recover it:
\begin{equation}
q_H\bigl(y^\star(x)\mid x\bigr)\approx 0,
\qquad
q_{S,a}\bigl(y^\star(x)\mid x\bigr)\approx 1.
\label{eq:reverse-captcha}
\end{equation}
Together with Equation~\eqref{eq:human-machine-gap}, this relationship gives $y_H^\dagger(x)\neq y^\star(x)$. 
Intuitively, the physical ground truth is difficult for humans to recover but comparatively easy for machines. Humans nevertheless pass reliably because the verifier accepts the human-reference answer 
$y_H^\dagger(x)$ 
rather than the physical ground truth. 

Equation~\eqref{eq:reverse-captcha} describes the idealized reverse-capability case and is stronger than strictly required for rejection. In practice, a solver need not recover $y^\star(x)$ exactly: an imperfect solver response is also rejected as long as it differs from $y_H^\dagger(x)$. 

As machines, particularly LALMs, become increasingly capable of perception and reasoning about objective properties of the physical world, it becomes harder to construct challenges that are easy for humans yet difficult for machines. Reverse-capability CAPTCHAs reverse this asymmetry by using challenges for which machines can readily recover the physical ground truth while humans reliably produce a different response due to perceptual limitations, so advances in machine capability can strengthen rather than erode the human--machine separation. 

\begin{figure*}[t]
    \centering
    \includegraphics[width=\textwidth]{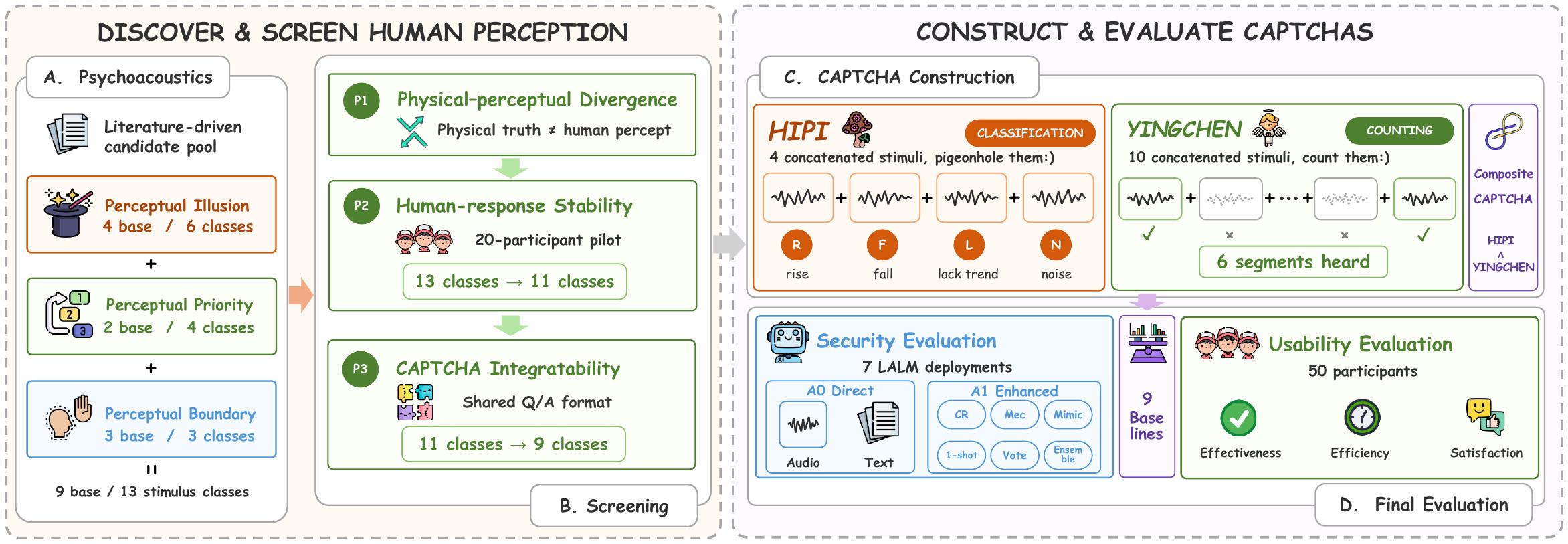}
    \caption{Overview of our reverse-capability audio CAPTCHA design and evaluation pipeline.}
    \label{fig:framework-overview}
\end{figure*}

\section{Threat Model}
\label{sec:threat-model}

\noindent\textbf{Adversary goals.}
Given a challenge $x$, adversaries select a solver $S$ and an attack strategy $a$. They attempt to use $S$ and $a$ to produce a response that is accepted by the verifier, i.e., $A(x,y_{S,a}(x))=1$. 

Such bypasses undermine the intended access-control role of CAPTCHAs by allowing automated systems to perform operations that would otherwise require successful human verification. Successful automated bypass can enable CAPTCHA-gated actions at scale, including bulk account registration, credential abuse, automated scraping, spam or fraudulent content distribution, and other forms of service abuse\cite{searles2023modern,guerar2021captcha}.

\noindent\textbf{Adversary capabilities.}
We consider an inference-only adversary that can integrate general-purpose LALMs into an automated solving pipeline. The adversary operates the models as black boxes and neither accesses their internal parameters nor modifies them through training or fine-tuning. It has access to the complete challenge audio and knows the response format. The adversary may use attacker-engineered prompts, CAPTCHA-specific design information or labeled examples, repeated sampling and response aggregation, and multiple LALMs. Section~\ref{sec:attack_strategy} instantiates these capabilities as A0 attacks, which exclude these enhancements, and A1 attacks, which permit one or more of them.

%% file: 04_empirical_study.tex
\section{Psychoacoustics-based Reverse-Capability Audio CAPTCHAs}

The reverse-capability framework establishes a general design principle: exploit stable human perceptual limitations such that the human-reference answer diverges from the physical ground truth. This principle admits multiple instantiations. In this work, we instantiate it through psychoacoustics by exploiting systematic divergences between objectively measurable acoustic properties and human auditory perception.

Psychoacoustics, however, encompasses a broad range of auditory phenomena, not all of which can be operationalized as stimulus classes that 
elicit the stable, composable human response patterns required for CAPTCHA construction. We therefore follow a two-stage design process, as summarized in \figurename~\ref{fig:framework-overview}. First, we conduct a pilot study to screen 13 candidate stimulus classes derived from nine base stimulus designs. 
Each candidate is grounded in a documented auditory phenomenon or perceptual limitation. Second, we combine the retained classes to construct two concrete reverse-capability audio CAPTCHAs.

\subsection{Pilot Study}
\label{sec:formative}

The pilot study aims to identify psychoacoustic stimulus classes suitable as building blocks for reverse-capability audio CAPTCHAs. We define three design principles that each candidate class should satisfy: 

\begin{itemize}[leftmargin=*]
\item \textbf{P1: Physical--perceptual divergence.}
A suitable stimulus class should exploit a human perceptual limitation that makes the predicted human response differ from the physical ground truth, while leaving that ground truth objectively recoverable from the waveform.

\item \textbf{P2: Human-response stability.}
{A suitable stimulus class should produce a reproducible perceptual outcome. For each stimulus condition, most listeners should consistently report the prespecified outcome, and this agreement should persist across independently generated instances of the same class.} 

\item \textbf{P3: CAPTCHA integratability.}
A suitable stimulus class should support diverse randomized instances and be composable with other classes under a shared task and response format, thereby yielding challenges with a sufficiently large joint answer space.

\end{itemize}

Together, P1 establishes the reverse-capability relationship, P2 supports reliable human verification, and P3 enables the construction of sufficiently diverse composite challenges.

\begin{table*}[t]
\centering
\caption{Candidate psychoacoustic stimulus classes and selection results.}
\label{tab:candidate-pool}
\footnotesize
\setlength{\tabcolsep}{3.0pt}
\renewcommand{\arraystretch}{1.1}
\begin{threeparttable}

\begin{tabularx}{\textwidth}{
@{}
>{\RaggedRight\arraybackslash}p{0.34\textwidth}
c
>{\RaggedRight\arraybackslash}X
>{\centering\arraybackslash}p{0.08\textwidth}
>{\centering\arraybackslash}p{0.10\textwidth}
>{\centering\arraybackslash}p{0.095\textwidth}
>{\RaggedRight\arraybackslash}p{0.098\textwidth}
@{}
}
\toprule
Stimulus class
& Category
& Task
& $y_H^\dagger$
& Target rate (\%)
& Integratable
& Decision \\
\midrule

\multicolumn{7}{@{}l}{
\textit{Shepard--Risset glissando family}~\cite{risset1969pitch}
} \\
\hspace{1em}Upward Shepard--Risset glissando
& \multirow{2}{*}{I}
& \multirow{2}{=}{\RaggedRight Dominant impression
(\texttt{R}/\texttt{F}/\texttt{L}/\texttt{N})}
& \texttt{R}
& 100
& \multirow{2}{*}{\cmark}
& \multirow{2}{=}{\RaggedRight Retain} \\
\hspace{1em}Downward Shepard--Risset glissando
&
&
& \texttt{F}
& 100
&
& \\
\addlinespace[2pt]

\multicolumn{7}{@{}l}{
\textit{Glide-disambiguated Shepard tritone family}~
\cite{frane2024shepardglide}
} \\
\hspace{1em}Upward glide-disambiguated Shepard tritone
& \multirow{2}{*}{I}
& \multirow{2}{=}{\RaggedRight Beginning/end pitch judgment
(\texttt{B}/\texttt{E}/\texttt{N})}
& \texttt{E}
& 100
& \multirow{2}{*}{\xmark}
& \multirow{2}{=}{\RaggedRight Excluded (P3)} \\
\hspace{1em}Downward glide-disambiguated Shepard tritone
&
&
& \texttt{B}
& 95
&
& \\
\addlinespace[2pt]

Missing fundamental~\cite{terhardt1974pitch}
& I
& Pitch match (\texttt{Y}/\texttt{N})
& \texttt{Y}
& 55
& \xmark
& Excluded (P2) \\

Zwicker tone~\cite{zwicker1964afterimage}
& I
& Post-noise tone (\texttt{Y}/\texttt{N})
& \texttt{Y}
& 15
& \xmark
& Excluded (P2) \\
\addlinespace[2pt]

\multicolumn{7}{@{}l}{
\textit{Moving spectral-ridge noise family}~\cite{horbach2018pitch}
} \\
\hspace{1em}Upward moving spectral-ridge noise
& \multirow{2}{*}{P}
& \multirow{2}{=}{\RaggedRight Dominant impression
(\texttt{R}/\texttt{F}/\texttt{L}/\texttt{N})}
& \texttt{N}
& 80
& \multirow{2}{*}{\cmark}
& \multirow{2}{=}{\RaggedRight Retain} \\
\hspace{1em}Downward moving spectral-ridge noise
&
&
& \texttt{N}
& 80
&
& \\
\addlinespace[2pt]

\multicolumn{7}{@{}l}{
\textit{Small-span high-frequency glide family}~\cite{sek1995frequency}
} \\
\hspace{1em}Small-span high-frequency upward glide
& \multirow{2}{*}{P}
& \multirow{2}{=}{\RaggedRight Dominant impression
(\texttt{R}/\texttt{F}/\texttt{L}/\texttt{N})}
& \texttt{L}
& 85
& \multirow{2}{*}{\cmark}
& \multirow{2}{=}{\RaggedRight Retain} \\
\hspace{1em}Small-span high-frequency downward glide
&
&
& \texttt{L}
& 80
&
& \\
\addlinespace[2pt]

Very-low-frequency tone~\cite{watanabe1990lowfreq}
& B
& Audible-segment count
& \texttt{0}
& 90
& \cmark
& Retain \\

Strongly attenuated sound~\cite{moore2012psychologyhearing}
& B
& Audible-segment count
& \texttt{0}
& 95
& \cmark
& Retain \\

Very-short silent gap~\cite{green1989temporalgaps}
& B
& Audible-segment count
& \texttt{1}
& 90
& \cmark
& Retain \\

\bottomrule
\end{tabularx}

\begin{tablenotes}
\item Note:
(i) I, P, and B denote perceptual illusion, perceptual priority,
and perceptual boundary, respectively.
(ii) The target rate is the percentage of human responses
matching $y_H^\dagger(x)$.
(iii) Italicized group rows denote stimulus families with directional
variants; each indented row represents a distinct stimulus class.
\end{tablenotes}

\end{threeparttable}
\end{table*}

\subsubsection{Candidate Pool}
\label{sec:candidate-pool}

To address P1, drawing on the psychoacoustic literature~\cite{bregman1990auditoryscene,moore2012psychologyhearing}, we compiled an initial pool of candidate stimulus classes for which human responses do not necessarily reflect properties objectively recoverable from the waveform. As summarized in \tablename~\ref{tab:candidate-pool}, we organize these stimulus classes into three categories according to how this physical--perceptual mismatch arises.

\noindent \textbf{Perceptual Illusion (I).}
The stimulus is clearly audible, but the auditory system reorganizes or supplements the available acoustic information, causing listeners to perceive a specific attribute or event, such as pitch, motion direction, or tone, that is absent from, ambiguous in, or inconsistent with the signal. 
We include the following four base stimulus designs, corresponding to six stimulus classes:

\begin{itemize}[leftmargin=*]
    \item \textit{Upward/Downward Shepard--Risset glissando.} 
    It combines several tones separated by octaves (frequencies differ by factors of two) and moves all of them upward or downward 
    simultaneously. As a component approaches one end of the frequency range, it gradually fades out, while a new component fades in at the other end. This replacement is difficult to notice, so the sound is physically recycled while listeners hear a tone that appears to rise or fall forever~\cite{shepard1964circularity,burns1981circularity,risset1969pitch,frane2024shepardglide}.  

\item \textit{Upward/Downward glide-disambiguated Shepard tritone.}
This stimulus consists of two Shepard tones separated by six semitones and connected by a continuous glide. Each Shepard tone combines the same note across several octaves, making its exact octave ambiguous. Without the glide, the second tone can be interpreted as either six semitones above or below the first. The pair may therefore sound ascending or descending depending on the particular notes and listener, an ambiguity known as the tritone paradox~\cite{deutsch1987tritone,malek2018pitchclass}. 
The connecting glide which rises (upward) or falls (downward) 
provides a clear cue to the direction of pitch motion and can thereby resolve this ambiguity~\cite{frane2024shepardglide}. 

\item \textit{Missing fundamental.}
A harmonic sound usually contains a fundamental frequency $f_0$ and higher-frequency components at integer multiples of $f_0$. Even if the component at $f_0$ is removed, listeners may still hear a pitch corresponding to it~\cite{terhardt1974pitch}. For example, a sound containing components at 200, 300, and 400~Hz, but not at 100~Hz, may nevertheless be heard as having a pitch of 100~Hz. The auditory system infers this pitch from the regular frequency spacing of the remaining components. 

\item \textit{Zwicker tone.}
A Zwicker tone occurs after a listener hears noise from which a narrow frequency band has been removed~\cite{zwicker1964afterimage}. When the noise stops, the listener may briefly hear a faint, decaying tone near the removed frequency band, even though the playback has ended and 
no actual tone is present. 
\end{itemize}

\noindent \textbf{Perceptual Priority (P).}
The overall sound is clearly audible and contains a physically measurable secondary feature, but the auditory system gives greater perceptual weight to a more dominant or global interpretation. The weaker feature therefore does not determine the listener's response. We include the following two base stimulus designs, corresponding to four stimulus classes: 

\begin{itemize}[leftmargin=*]

{
\item \textit{Upward/downward moving spectral-ridge noise.}
This stimulus consists of broadband noise whose spectral energy is shaped into a broad region that moves continuously upward or downward in frequency over time. This motion forms a sloping ridge in the spectrogram, from which signal analysis can recover the direction of movement. However, broad noise-like signals provide relatively weak pitch and frequency-motion cues, so listeners may not reliably identify the direction of the moving ridge and may instead perceive the stimulus primarily as noise~\cite{horbach2018pitch,faulkner2000salience}.
}

\item \textit{Small-span high-frequency upward/downward glide.}
This stimulus is a high-pitched tone whose frequency continuously increases (upward) or decreases (downward) 
by a small amount. Although the frequency change can be measured directly from the waveform, it may be too small for listeners to identify, particularly at frequencies above 4~kHz, where sensitivity to the direction of frequency change deteriorates. 
Listeners may therefore hear an approximately steady tone or notice a change without reliably determining whether it rises or falls~\cite{sek1995frequency}. 
\end{itemize}

\noindent \textbf{Perceptual Boundary (B).}
A physical sound or event falls below a human detection or resolution threshold.  
Unlike perceptual priority, where the target feature is overshadowed by a dominant percept, the target event itself does not produce a reliably distinguishable auditory sensation. We consider the following three stimulus classes:

\begin{itemize}[leftmargin=*]
    \item \textit{Very-low-frequency tone.}
Human hearing becomes much less sensitive as frequency approaches the lower end of the audible range. A very-low-frequency tone must therefore be played at a much higher sound-pressure level than a mid-frequency tone to be heard. If its playback level remains below the hearing threshold for that frequency, the tone is physically present in the waveform but may be inaudible to listeners~\cite{watanabe1990lowfreq}.

\item \textit{Strongly attenuated sound.}
Attenuation reduces the amplitude of a sound without removing it from the digital waveform. If the attenuated sound produces a level below the listener's hearing threshold during playback, it becomes inaudible even though its acoustic structure remains measurable from the waveform~\cite{moore2012psychologyhearing}.

\item \textit{Very-short silent gap.}
This stimulus contains two sound segments separated by a physically silent interval. Because human hearing has limited temporal resolution, a gap lasting only a few milliseconds may fall below the listener's gap-detection threshold. The listener may consequently hear one continuous sound rather than two segments separated by silence~\cite{phillips1999gap,green1989temporalgaps}.
\end{itemize}

The candidate pool comprises 9 base stimulus designs. Four of these designs, Shepard--Risset glissando, glide-disambiguated Shepard tritone, small-span high-frequency glide, and {moving spectral-ridge noise}, have both upward and downward variants, yielding 13 stimulus classes in total. 

\subsubsection{Candidate Selection by P2}
\label{sec:pilot-design}

We conduct a questionnaire-based listening study through CloudResearch Connect~\cite{cloudresearch2026connect}, an online professional survey and participant-recruitment platform, to determine which candidate stimulus classes satisfy P2. The study was approved by the Institutional Review Board (IRB) at our institution. 

\noindent\textbf{Audio stimulus generation.}
{Each stimulus class has a set of controllable generation parameters, such as waveform content, phase, noise realization, and initial frequency.} For each of the 13 classes, we independently generate 20 audio stimuli {by varying its class-specific parameters,} yielding $13 \times 20 = 260$ stimuli in total. {Appendix~\ref{app:formative-materials} provides further details on the generation process.}

\noindent\textbf{Tasks.}
The task depends on the stimulus class. For the six directional stimulus classes, namely the upward and downward variants of the Shepard--Risset glissando, small-span high-frequency glide, and {moving spectral-ridge noise}, participants classify the dominant auditory impression as rising (\texttt{R}), falling (\texttt{F}), lacking a clear trend (\texttt{L}), or noise (\texttt{N}). For the two glide-disambiguated Shepard tritone classes, they indicate whether the beginning (\texttt{B}) or ending (\texttt{E}) is higher in pitch, or whether the direction is unclear (\texttt{N}). 
For the missing-fundamental class, participants judge whether a test stimulus with its fundamental component removed has the same perceived pitch as a corresponding reference stimulus with that component retained. For the Zwicker-tone class, they report whether they hear a tone after the noise ends. Both tasks use binary responses (\texttt{Y} or \texttt{N}). For the three perceptual-boundary stimulus classes, participants report the number of audible sound segments.

\noindent \textbf{Participants.} 
We recruit 20 participants. For each stimulus class, we assign its 20 independently generated instances one-to-one to the participants. Each participant therefore listens to and responds to 13 stimuli with one from each class, and no stimulus instance is shared across participants.  
To ensure response quality, using the platform's built-in prescreening filters, we restrict eligibility to adults who report English as their native language and are located in English-speaking countries. 
We also require participants to use headphones or earphones and complete the study in a quiet environment. Before proceeding, they must confirm that both requirements are satisfied via a mandatory checkbox. 

\noindent\textbf{Selection.}
For each stimulus $x$, we prespecify a human-reference answer $y_H^\dagger(x)$. A stimulus class passes P2 only if at least 80\% of the responses to its instances match their corresponding human-reference answers.

As summarized in \tablename~\ref{tab:candidate-pool}, 11 of the 13 candidate stimulus classes pass P2, with target-response rates ranging from 80\% to 100\%. The missing-fundamental and Zwicker-tone classes fall below the threshold, with target-response rates of 55\% and 15\%, respectively. We therefore exclude both classes because the observed human responses do not show sufficient agreement with their prespecified human-reference answers.

\subsubsection{Candidate Selection by P3}
\label{sec:rq2-results}

For the retained stimulus classes, we examine whether a class can be combined with other candidates under a common question and answer format, allowing multiple instances to form a CAPTCHA with a larger joint answer space. 

Nine stimulus classes pass the examination and form two distinct compatible groups. The Shepard--Risset glissando, small-span high-frequency glide, and {moving spectral-ridge noise}, each with upward and downward variants, share a four-way categorical answer space: \emph{rising}, \emph{falling}, \emph{lacking clear overall trend}, and \emph{noise}. 
The very-low-frequency tone, strongly attenuated sound, and very-short silent gap can all be formulated as counting tasks and therefore share a numerical answer format. Within each group,
generated instances can be combined without changing the question or response semantics.

Both upward and downward variants of the glide-disambiguated Shepard tritone fail to pass the examination. They support only a comparison task with three possible responses: \emph {beginning tone higher}, \emph{end tone higher}, and \emph{unclear}. This answer space is both smaller and incompatible with the categorical and counting formats above. Consequently, these two variants cannot be combined with the other candidates without changing the task, while using them alone provides insufficient answer diversity.

%% file: 05_methodology.tex
\subsection{Audio CAPTCHA Design}
\label{sec:methodology}

Building on the psychoacoustic stimulus classes retained in Section~\ref{sec:formative}, we develop two audio CAPTCHAs: \textbf{\textsc{Hipi}} (\textbf{H}uman \textbf{I}nstinct \textbf{P}igeonholes \textbf{I}t), which uses a classification task, and \textbf{\textsc{Yingchen}} (\textbf{Y}our \textbf{I}nner \textbf{N}umber \textbf{G}remlin \textbf{C}ounts \textbf{H}eard \textbf{E}vents \textbf{N}aturally), which uses a counting task. Each combines multiple compatible stimulus classes. 
For both CAPTCHAs, the verifier’s reference answers follow the human-reference mappings summarized in \tablename~\ref{tab:physical-accepted-responses}, with the psychoacoustic-class mappings validated in the pilot study.

\subsubsection{Classification-based Audio CAPTCHA: \textsc{Hipi}}
\label{sec:hipi}
\mbox{}\par
\smallskip
\noindent\textbf{Task format.}
\textsc{Hipi} combines 
six 
psychoacoustic stimulus classes: 
upward and downward variants of Shepard--Risset glissandi, small-span high-frequency glides, and moving spectral-ridge noise. 
Each challenge $x_H$ presents an ordered sequence of $K$ {independently generated} audio stimuli, each belonging to one of these classes. For each stimulus $s_i$, the task is to assign one of four labels according to its dominant auditory impression: \texttt{R} for a clear rising trend, \texttt{F} for a clear falling trend, \texttt{L} when no clear overall trend is perceived, or \texttt{N} when noise is the dominant impression. The response to the challenge is the corresponding ordered label sequence ${y}=(y_1,\ldots,y_K)$. For example, when $K=4$, the sequence may be represented as \texttt{RFLN}.

\noindent\textbf{Verification.}
For each stimulus $s_i$, the verifier uses the human-reference label $y_H^\dagger(s_i)$ validated in the pilot study. These labels form the reference sequence
${y}^{\dagger}=(y_H^\dagger(s_1),\ldots,y_H^\dagger(s_K))$.

Given a raw submitted answer $r_H$, a parser $\operatorname{norm}_H(\cdot)$ removes whitespace and permitted separators, converts the remaining symbols to uppercase, and requires exactly $K$ symbols from ${\texttt{R}, \texttt{F}, \texttt{L}, \texttt{N}}$. For a valid submission, it returns the resulting label sequence; otherwise, it returns $\bot$. {Further parsing 
details are provided in Appendix~\ref{app:parsing}.} Verification is therefore
\begin{equation}
A_H(x_H,r_H)
=
\mathbb{I}\left[
\operatorname{norm}_H(r_H)={y}^{\dagger}
\right].
\label{eq:hipi-verifier}
\end{equation}
Thus, every parsed symbol must match its expected label in presentation order. 

\noindent\textbf{Challenge implementation.}
Each \textsc{Hipi} challenge contains $K=4$ stimuli. We choose $K=4$ as a practical trade-off between usability and security. A larger $K$ would lengthen the challenge and increase the user's listening, memory, and response burden, whereas a smaller $K$ would reduce the joint answer space and weaken resistance to guessing. For each stimulus, the generator independently and uniformly samples one of the 
six 
\textsc{Hipi} stimulus classes. Conditional on the selected class, it samples the {class-specific synthesis parameters from their prespecified distributions (Appendix~\ref{app:formative-materials})} and renders the corresponding stimulus. 

The rendered stimuli are assembled into a single audio clip. We prepend a spoken instruction explaining the task and the four response labels (see Appendix~\ref{app:audio-assembly} for details). A digitally silent interval with a duration of 450 milliseconds (ms) separates the instruction from the first stimulus and every pair of adjacent stimuli, with no temporal overlap. 

The interface presents the resulting clip through a single playback control and collects the $K$-character label sequence in a free-response text field. Presenting one composite clip avoids requiring users to manually initiate $K$ separate playbacks. We use free response rather than a single multiple-choice list of complete answers because such a list would contain up to $4^K$ options. 

\begin{table}[t]
    \centering
    \caption{Physical ground truth and human-reference answer for each stimulus class.
    The arrows $\uparrow$ and $\downarrow$ denote upward and downward variants, respectively.}
    \label{tab:physical-accepted-responses}
    \small
    \setlength{\tabcolsep}{5pt}
    \renewcommand{\arraystretch}{1.12}
    \begin{tabularx}{\columnwidth}{
        @{}
        X
        c
        c
        @{}
    }
        \toprule
        Stimulus class & $y^\star(x)$ & $y_H^\dagger(x)$ \\
        \midrule

        \multicolumn{3}{@{}l}{\textbf{\textsc{Hipi}}} \\
        Shepard--Risset glissando
        & \texttt{L}
        & $\uparrow$: \texttt{R}, $\downarrow$: \texttt{F} \\

        Small-span high-frequency glide
        & $\uparrow$: \texttt{R}, $\downarrow$: \texttt{F}
        & \texttt{L} \\

        {Moving spectral-ridge noise}
        & $\uparrow$: \texttt{R}, $\downarrow$: \texttt{F}
        & \texttt{N} \\

        \midrule
        \multicolumn{3}{@{}l}{\textbf{\textsc{Yingchen}}} \\
        Audible control
        & $1$ & $1$ \\
        Very-low-frequency tone
        & $1$ & $0$ \\
        Strongly attenuated sound
        & $1$ & $0$ \\
        Very-short silent gap
        & $2$ & $1$ \\

        \bottomrule
    \end{tabularx}
\end{table}

\subsubsection{Counting-based Audio CAPTCHA: \textsc{Yingchen}}
\label{sec:yingchen-mechanism}

\noindent\textbf{Task format.}
\textsc{Yingchen} combines three psychoacoustic stimulus classes retained from the pilot study---very-low-frequency tones (\texttt{LF}), strongly attenuated sounds (\texttt{SA}), and very-short silent gaps (\texttt{SS})---with a newly introduced audible control class (\texttt{A}) containing clearly audible medium- or narrow-band noise. Each challenge $x_{Y}$ presents an ordered sequence of $K$ audio stimuli, each belonging to one of these classes. Rather than classifying individual stimuli, the user reports the total number of audible sound segments perceived across the entire sequence. The response therefore should be a single integer $\hat{r}$. 

\noindent\textbf{Verification.}
For each stimulus $s_i$, the verifier assigns a human-reference audible-segment contribution $y_H^\dagger(s_i)\in\{0,1\}$. For the psychoacoustic stimulus classes, these contributions are validated in the pilot study; for the audible control class \texttt{A}, the contribution is set to one by construction. Thus, an \texttt{A} or \texttt{SS} stimulus contributes one perceived segment, whereas an \texttt{LF} or \texttt{SA} stimulus contributes none. The human-reference answer is therefore
\begin{equation}
r=\sum_{i=1}^{K}y_H^\dagger(s_i)=n_A+n_{SS},
\label{eq:yingchen-reference}
\end{equation}
where $n_A$ and $n_{SS}$ denote the numbers of \texttt{A} and \texttt{SS} stimuli, respectively.

Given a raw submitted answer $\hat{r}$, a parser $\operatorname{norm}_Y(\cdot)$ accepts exactly one unsigned decimal integer and returns its numerical value. Responses containing explanations, additional numbers, or ambiguous content are mapped to $\bot$. Verification is therefore
\begin{equation}
A_Y(x_Y,\hat{r})
=
\mathbb{I}\left[
\operatorname{norm}_Y(\hat{r})=r
\right].
\label{eq:yingchen-verifier}
\end{equation}

\noindent\textbf{Challenge implementation.}
Each \textsc{Yingchen} challenge contains $K=10$ stimuli. Unlike \textsc{Hipi}, the stimulus classes are not sampled independently. Instead, the generator first randomly selects a target answer \(r \in \{2,\ldots,8\}\) and then samples a valid class allocation satisfying
\begin{equation}
n_A+n_{LF}+n_{SA}+n_{SS}=K,
\qquad
n_A+n_{SS}=r.
\label{eq:yingchen-allocation}
\end{equation}
It then randomizes the order of the assigned classes and renders each stimulus using its corresponding class-specific parameter distributions.

The audio construction and presentation otherwise follow those of \textsc{Hipi}, including the spoken instruction, silent separators 
(except that the duration of each inter-stimulus silent interval is independently sampled from 200 to 300~ms), 
composite-clip presentation, and single playback control. The instruction is adapted to describe the counting task, and the interface collects a single numerical answer in the free-response field rather than a $K$-character label sequence.

\subsubsection{Composition of the Audio CAPTCHAs}
\label{sec:verifier-composition}

In addition to being deployed independently, \textsc{Hipi} and \textsc{Yingchen} can be composed into a composite challenge. Given a pair of challenges $(x_H,x_Y)$ and their submitted responses $(r_H,r_Y)$, the composite verifier accepts if and only if both component verifiers accept:
\begin{equation}
\begin{aligned}
&A_{\mathrm{comp}}((x_H,x_Y),(r_H,r_Y))\\
&\qquad =
A_H(x_H,r_H)\land A_Y(x_Y,r_Y).
\end{aligned}
\end{equation}
Consequently, failure on either component causes the composite challenge to be rejected.

%% file: 06_evaluation.tex
\section{Evaluation}
\label{sec:evaluation}

\subsection{Experimental Setup}

\subsubsection{Solvers}
We consider two types of solvers.

\noindent\textbf{LALM-based Solvers.}
We evaluate seven representative general-purpose LALMs spanning multiple model families and providers. Four are accessed through APIs: gpt-realtime-2~\cite{openai2026gptrealtime2}, gpt-audio-1.5~\cite{openai2026gptaudio15}, qwen3.5-omni-plus~\cite{qwen2026qwen35omni}, and qwen3.5-omni-flash~\cite{qwen2026qwen35omni}. The remaining three are open-weight models deployed locally: kimi-audio-7b~\cite{kimi2025audio}, nemotron-3-nano-omni~\cite{nvidia2026nemotron3omni}, and qwen2-audio-7b~\cite{chu2024qwen2audio}.

\noindent\textbf{Human-based Solvers.}
We additionally conduct a human study to evaluate human solving performance and usability on \textsc{Hipi} and \textsc{Yingchen}.

\subsubsection{Dataset}

The main evaluation set comprises 100 \textsc{Hipi} challenges and 100 \textsc{Yingchen} challenges generated using the procedure described in Section~\ref{sec:methodology}. All constituent audio stimuli are newly generated and disjoint from those used in the pilot study. Each challenge contains the complete audio clip and its corresponding human-reference answer.

For the A1-One-shot attack (Section~\ref{sec:attack_strategy}), we generate one additional \textsc{Hipi} challenge and one additional \textsc{Yingchen} challenge. Their constituent stimuli are disjoint from both the pilot-study and main-evaluation sets. 

To evaluate the composition of the two CAPTCHAs, we establish a random one-to-one pairing between the 100 \textsc{Hipi} challenges and the 100 \textsc{Yingchen} challenges, yielding 100 composite challenges.

\subsubsection{Baselines}
We compare our audio CAPTCHAs with 
nine 
representative prior audio CAPTCHA schemes, covering both transcription-based and reasoning-based designs. The transcription-based baselines include aaeCAPTCHA~\cite{hossen2022aaecaptcha}, Yeehaw Junction~\cite{abdullah2023yeehaw}, Control~\cite{fanelle2020accessibility}, and Pauses~\cite{fanelle2020accessibility}, while the reasoning-based baselines include Meutzner et al.~\cite{meutzner2017improved}, Volcano~\cite{lee2024advcaptcha}, Math, Character~\cite{fanelle2020accessibility}, 
and Categories~\cite{fanelle2020accessibility}. 
We reproduce these schemes following their official designs and generate 100 challenges for each scheme. For each scheme, we additionally generate one instance that does not overlap with the evaluation challenges for use in the A1-One-shot attack (cf.~Section~\ref{sec:attack_strategy}). 
We exclude the concurrent IllusionAudio~\cite{ding2026robustcaptchausingaudio}
because its official implementation is not publicly available, and faithful reproduction is not trivial. 

\subsubsection{Metrics}
\label{sec:security-usability-goals}
We consider two dimensions.

\noindent\textbf{Security.}
We measure security by the rate at which machine-generated responses are accepted by the verifier; lower values indicate stronger security. Let $\mathcal{S}$ and $\mathcal{A}$ denote the sets of solvers and attack strategies, respectively. We report machine acceptance rate (MAR) at three aggregation levels.

\begin{itemize}[leftmargin=*]
\item \textbf{Configuration-level MAR ($\mathrm{MAR}_{S,a}$).}
For solver $S\in\mathcal{S}$ under attack strategy $a\in\mathcal{A}$, $\mathrm{MAR}_{S,a}$ is the expectation that its response is accepted:
\begin{equation}
\mathrm{MAR}_{S,a}
=
\mathbb{E}_{\substack{x\sim\mathcal{D},\
y_{S,a}\sim q_{S,a}(\cdot\mid x)}}
\mathbb{I}\left[A(x,y_{S,a})=1\right].
\end{equation}
Given an evaluation dataset
$\mathcal{X}=\{x_i\}_{i=1}^{N}$ and one observed response
$y_{S,a}^{i}$ for each challenge, we estimate this expectation as
\begin{equation}
\mathrm{MAR}_{S,a}
=
\frac{1}{N}
\sum_{i=1}^{N}
\mathbb{I}\left[A(x_i,y_{S,a}^{i})=1\right].
\end{equation}
Thus, the empirical MAR is the proportion of evaluated challenges for which the solver's response is accepted. 

\item \textbf{Strategy-level maximum MAR (${\mathrm{MAR}}_{a}$).}
For each fixed attack strategy $a$, we take the maximum MAR across solvers:
\begin{equation}
     {\mathrm{MAR}}_{a}
    =
    \max_{\substack{S\in\mathcal{S}}}
    \mathrm{MAR}_{S,a}.
\end{equation}
This provides a strategy-level summary of machine acceptance without focusing on any single solver.

\item \textbf{Solver-level maximum MAR ($\mathrm{MAR}_{S}$).}
For each fixed solver $S$, we take the maximum MAR across attack strategies: 
\begin{equation}
    \mathrm{MAR}_{S}
    =
    \max_{\substack{a\in\mathcal{A}}}
    \mathrm{MAR}_{S,a}.
\end{equation}
This represents the strongest attack under a specific solver. 

\end{itemize}

A1-Ensemble (cf.~Section~\ref{sec:attack_strategy}) is a cross-model strategy rather than a solver-specific configuration. It therefore has no $\mathrm{MAR}_{S,a}$; we report its aggregate MAR directly as $\mathrm{MAR}_{a}$ and exclude it when computing $\mathrm{MAR}_{S}$.

\noindent\textbf{Usability.}
We evaluate usability through a human study following the effectiveness, efficiency, and satisfaction dimensions of ISO~9241-11~\cite{iso9241_11_2018}. We measure effectiveness using the human acceptance rate (HAR), efficiency using completion time and replay rate, and satisfaction using task-level satisfaction ratings.

HAR is the probability that a human response is accepted by the verifier:
\begin{equation}
\mathrm{HAR}
=
\mathbb{E}_{\substack{x\sim\mathcal{D},\
y_H\sim q_H(\cdot\mid x)}}
\mathbb{I}\left[A(x,y_H)=1\right].
\end{equation}
Let
$\mathcal{O}={(x_j,y_H^{j})}_{j=1}^{M}$
denote the $M$ observed human challenge--response pairs collected in the study. We estimate HAR as 
\begin{equation}
\mathrm{HAR}
=
\frac{1}{M}
\sum_{j=1}^{M}
\mathbb{I}\left[A(x_j,y_H^{j})=1\right].
\end{equation}
Thus, empirical HAR is the proportion of observed human responses accepted by the verifier. Further details of the human study are provided in Section~\ref{sec:rq4-usability}.

\subsubsection{Attack Strategies}\label{sec:attack_strategy}

Following the attacker capabilities defined in Section~\ref{sec:threat-model}, we evaluate two capability tiers in our experiments. They are distinguished by four forms of attack enhancement: (i) attacker-engineered prompting; (ii) CAPTCHA-specific knowledge or labeled examples; (iii) repeated inference and response aggregation; and (iv) access to multiple solver models. A0 attacks exclude all four enhancements, whereas A1 attacks permit one or more of them. 

\noindent\textbf{Attacker A0: Baseline attacker.}
A0 performs a single inference using one general-purpose LALM with fixed decoding settings. It has access only to the challenge audio and information directly derived from it. 
We develop two A0 attackers: 

\begin{itemize}[leftmargin=*]
\item \textit{A0-Audio:} The complete challenge audio, including its spoken task instruction, is provided to the model without an additional text prompt. 

\item \textit{A0-Text:} The complete challenge audio is provided together with a verbatim text transcript of the task instruction. No attacker-engineered guidance or CAPTCHA-specific information is included. In practice, an attacker could extract the instruction from the audio using ASR. We instead provide the ground-truth transcript, eliminating errors from this preprocessing step and thereby giving the attacker a stronger input condition.  

\end{itemize}

\noindent\textbf{Attacker A1: Enhanced attacker.}
A1 remains inference-only but relaxes one or more of the A0 restrictions. The following attacks separately evaluate prompt engineering, CAPTCHA-adaptive information, repeated sampling, and cross-model aggregation:

\begin{itemize}[leftmargin=*]
\item \textit{A1-CR:} The attacker adds a reasoning-oriented prompt instructing the model to analyze the challenge carefully before returning its final answer.

\item \textit{A1-Mec:} The attacker understands the CAPTCHA mechanism and provides information about its design principles to help the model better understand and solve the challenge. 

\item \textit{A1-Mimic:} The attacker informs the model of the CAPTCHA's reverse-capability design and instructs it to predict the auditory judgment of an ordinary human listener rather than directly report the physical properties of the signal.

\item \textit{A1-One-shot:} The attacker provides one complete CAPTCHA example together with its verifier-accepted response before presenting the audio challenge.

\item \textit{A1-Vote:} The attacker independently samples multiple responses from the same LALM and submits the modal (most frequent) response.

\item \textit{A1-Ensemble:} The attacker queries multiple distinct LALMs using the same challenge and submits the modal response across models.

\end{itemize}

A1-CR incorporates prompt engineering, whereas A1-Mec, A1-Mimic, and A1-One-shot additionally exploit CAPTCHA-specific knowledge and are therefore adaptive to the proposed design. A1-Vote and A1-Ensemble further evaluate repeated-sampling inference and cross-model aggregation, respectively. 
We use a budget of 7 model invocations per challenge: A1-Vote invokes each solver seven times, whereas A1-Ensemble invokes each of the seven solvers once; all other strategies exploit a single invocation. 

\subsection{Security against LALM-based Solvers}
\label{sec:rq3-security}

\begin{figure*}[t]
    \centering
    \includegraphics[width=\textwidth]{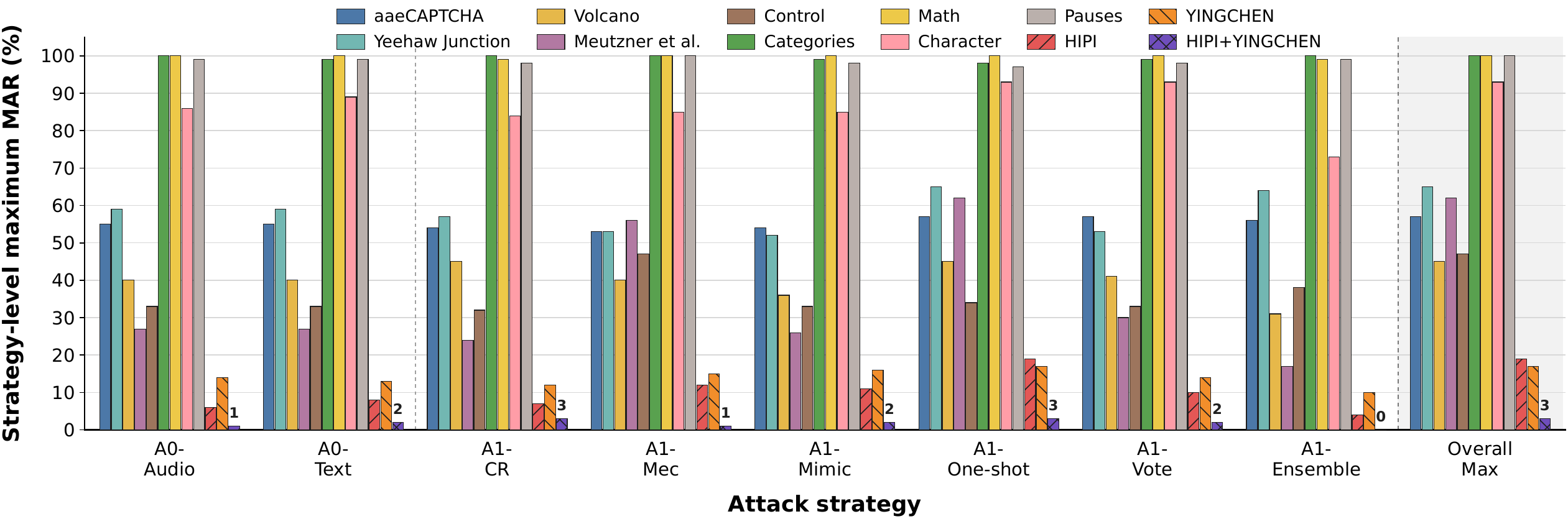}
    \vspace*{-4mm}
    \caption{Comparison of strategy-level maximum MARs among \textsc{Hipi}, \textsc{Yingchen}, \textsc{Hipi}+\textsc{Yingchen}, and baselines.}
    \label{fig:solver-averaged-mar}
\end{figure*}

This section first evaluates and compares the resilience of the proposed and baseline audio CAPTCHAs against LALM-based solvers (Section~\ref{sec:overall_per}), and then evaluates the impact of solvers and attack strategies on solving \ao and \at (Section~\ref{sec:impact_solver_strategy}). 
More details of the evaluation setup are provided in Appendix~\ref{app:machine_materials}. 

\subsubsection{Overall performance}\label{sec:overall_per}
We compare the strategy-level maximum MARs across solvers for different attack strategies among \textsc{Hipi}, \textsc{Yingchen}, \textsc{Hipi}+\textsc{Yingchen}, and baselines. 
The results are shown in \figurename~\ref{fig:solver-averaged-mar}, where lower 
MARs indicate stronger resistance to LALM-based attacks. 

\noindent\textbf{Comparison with baselines.}
All three proposed audio CAPTCHAs consistently outperform the baselines. Under the two A0 attacks, the baseline MARs range from 27\% to 100\%, whereas \textsc{Hipi}, \textsc{Yingchen}, and their composition achieve 6--8\%, 13--14\%, and 1--2\%, respectively. The advantage persists under A1 attacks: the baseline MARs range from 17\% to 100\%, compared with 4--19\% for \textsc{Hipi}, 10--17\% for \textsc{Yingchen}, and 0--3\% for the composition. 
In the overall-maximum summary, which takes the maximum MAR over all attack strategies, the baselines obtain 45--100\%, while \textsc{Hipi}, \textsc{Yingchen}, and their composition remain at 19\%, 17\%, and 3\%, respectively. 

{In addition, we find that the baseline Meutzner is particularly sensitive to CAPTCHA-specific adaptation: mechanism-aware and one-shot prompting increase its MAR from 27\% under A0 to 56\% and 62\%, respectively. In contrast, both proposed tasks remain below 20\% under all attack strategies. These results suggest that CAPTCHA-specific knowledge and demonstrations can help automated solvers recover the response rules of some prior designs more readily than the human-oriented responses required by the proposed reverse-capability tasks.}

These results indicate that reverse-capability verification provides stronger resistance to current LALMs. 

\noindent\textbf{Comparison between \ao and \at.}
\textsc{Hipi} achieves a lower MAR than \textsc{Yingchen} under seven of the eight individual attack strategies, indicating stronger resistance under most attack conditions. The only exception is A1-One-shot, under which the MAR of \textsc{Hipi} increases to 19\%, slightly exceeding the 17\% of \textsc{Yingchen}. 
A plausible explanation is that \textsc{Hipi}'s four-way perceptual classification task is less immediately interpretable than \textsc{Yingchen}'s counting task, allowing the one-shot example to more effectively clarify the intended judgment and label mapping. 
Consequently, although \textsc{Hipi} generally yields lower MARs, \textsc{Yingchen} has a slightly lower overall maximum MAR because it is less affected by the one-shot adaptation.

\noindent\textbf{Benefit of composition.}
Conjunctively composing the two CAPTCHAs further strengthens security. The composite MAR ranges from 0\% to 3\% and is lower than the MAR of either single CAPTCHA under every attack strategy. Its overall maximum MAR is only 3\%, representing absolute reductions of 16 and 14 percentage points relative to \textsc{Hipi} and \textsc{Yingchen}, respectively. These results show that an attacker capable of solving one task does not necessarily solve the other and that requiring both task-specific responses to be accepted substantially reduces overall machine acceptance.

\subsubsection{Impact of Solvers and Attack Strategies}\label{sec:impact_solver_strategy}

\begin{table*}[t]
    \centering
    \caption{Machine acceptance rates (MAR; \%). Each triplet reports
    \textsc{Hipi} / \textsc{Yingchen} / their composite.}
    \label{tab:a0-a1-mar-full}

    \setlength{\tabcolsep}{2.5pt}
    \renewcommand{\arraystretch}{1.00}

    \resizebox{1\textwidth}{!}{%
    \begin{tabular}{@{}l*{9}{c}@{}}
        \toprule
        \multirow{2}{*}{\diagbox{Solver}{Strategy}}
        & \multicolumn{2}{c}{A0}
        & \multicolumn{7}{c}{A1} \\
        \cmidrule(lr){2-3}
        \cmidrule(lr){4-10}

        & Audio
        & Text
        & CR
        & Mec
        & Mimic
        & One-shot
        & Vote
        & Ensemble
        & $\mathrm{MAR}_{S}$ \\
        \midrule

        gpt-realtime-2
        & \MARmy{3}{12}{0}
        & \MARmy{7}{10}{2}
        & \MARmy{3}{8}{0}
        & \MARmy{3}{10}{0}
        & \MARmy{4}{14}{0}
        & \MARmy{10}{15}{3}
        & \MARmy{6}{12}{0}
        & \multirow{7}{*}{\MARmy{4}{10}{0}}
        & \bfseries\MARmy{10}{15}{3} \\

        gpt-audio-1.5
        & \MARmy{0}{0}{0}
        & \MARmy{3}{0}{0}
        & \MARmy{0}{0}{0}
        & \MARmy{0}{1}{0}
        & \MARmy{0}{12}{0}
        & \MARmy{2}{16}{0}
        & \MARmy{0}{0}{0}
        &
        & \bfseries\MARmy{3}{16}{0} \\

        qwen3.5-omni-plus
        & \MARmy{6}{10}{1}
        & \MARmy{8}{11}{2}
        & \MARmy{7}{3}{3}
        & \MARmy{12}{9}{0}
        & \MARmy{11}{11}{2}
        & \MARmy{19}{7}{1}
        & \MARmy{10}{12}{2}
        &
        & \bfseries\MARmy{19}{12}{3} \\

        qwen3.5-omni-flash
        & \MARmy{1}{12}{0}
        & \MARmy{0}{13}{0}
        & \MARmy{0}{11}{0}
        & \MARmy{0}{15}{0}
        & \MARmy{0}{16}{0}
        & \MARmy{8}{13}{0}
        & \MARmy{1}{14}{1}
        &
        & \bfseries\MARmy{8}{16}{1} \\

        kimi-audio-7b
        & \MARmy{0}{14}{0}
        & \MARmy{2}{10}{1}
        & \MARmy{2}{12}{1}
        & \MARmy{1}{11}{1}
        & \MARmy{1}{11}{0}
        & \MARmy{0}{17}{0}
        & \MARmy{1}{13}{0}
        &
        & \bfseries\MARmy{2}{17}{1} \\

        nemotron-3-nano-omni
        & \MARmy{0}{8}{0}
        & \MARmy{0}{6}{0}
        & \MARmy{0}{6}{0}
        & \MARmy{1}{6}{0}
        & \MARmy{0}{1}{0}
        & \MARmy{1}{12}{0}
        & \MARmy{0}{10}{0}
        &
        & \bfseries\MARmy{1}{12}{0} \\

        qwen2-audio-7b
        & \MARmy{0}{0}{0}
        & \MARmy{0}{0}{0}
        & \MARmy{0}{10}{0}
        & \MARmy{0}{6}{0}
        & \MARmy{0}{3}{0}
        & \MARmy{0}{11}{0}
        & \MARmy{0}{9}{0}
        &
        & \bfseries\MARmy{0}{11}{0} \\

        \midrule
        \bfseries $\mathrm{MAR}_{a}$
        & \bfseries\MARmy{6}{14}{1}
        & \bfseries\MARmy{8}{13}{2}
        & \bfseries\MARmy{7}{12}{3}
        & \bfseries\MARmy{12}{15}{1}
        & \bfseries\MARmy{11}{16}{2}
        & \bfseries\MARmy{19}{17}{3}
        & \bfseries\MARmy{10}{14}{2}
        & \bfseries\MARmy{4}{10}{0}
        & \bfseries\MARmy{19}{17}{3} \\

        \bottomrule
    \end{tabular}%
    }
\end{table*}

We report the configuration-level MAR for each solver-strategy combination, the strategy-level maximum MAR across all solvers, and the solver-level maximum MAR across all strategies. The results are shown in \tablename~\ref{tab:a0-a1-mar-full}. 

\noindent\textbf{Comparison across solvers.}
As shown in \tablename~\ref{tab:a0-a1-mar-full}, attack effectiveness is strongly solver- and task-dependent. qwen3.5-omni-plus is the strongest solver against \textsc{Hipi}, reaching 19\% under A1-One-shot, whereas kimi-audio-7b is strongest against \textsc{Yingchen}, reaching 17\% under the same strategy. Notably, kimi-audio-7b reaches only 2\% on \textsc{Hipi}, while several other solvers similarly remain near zero on \textsc{Hipi} but achieve nontrivial acceptance on \textsc{Yingchen}. Thus, no solver dominates both tasks. Moreover, the composite MAR never exceeds 3\% for any solver across the evaluated attack strategies, indicating that a solver's strength on one task does not reliably transfer to the composite CAPTCHA.

\noindent\textbf{A0 versus A1.}
Inference-time adaptation improves the strongest attack across solvers, particularly against \textsc{Hipi}. Across the two A0 strategies, the maximum MARs are 8\%, 14\%, and 2\% for \textsc{Hipi}, \textsc{Yingchen}, and their composition, respectively. The corresponding maxima across A1 rise to 19\%, 17\%, and 3\%. A1 therefore provides a substantial gain against \textsc{Hipi} but only modest gains against \textsc{Yingchen} and the composition. 
Relative to A0-Audio, every other strategy except A1-Ensemble yields a higher maximum MAR on \textsc{Hipi}, whereas the effects on \textsc{Yingchen} are mixed. One possible explanation is that \textsc{Hipi} is more difficult for LALMs and has a lower baseline MAR, leaving greater room for adaptation. A1-Ensemble instead reduces MAR in all three settings, likely because the incorrect responses of most constituent solvers outvote the correct responses of the few successful ones. 

\noindent\textbf{A0-Audio versus A0-Text.}
Providing the spoken task instruction additionally as text has only a small and non-uniform effect. 
The strategy-level maximum MAR increases from 6\% to 8\% for \textsc{Hipi}, decreases from 14\% to 13\% for \textsc{Yingchen}, and remains at most 2\% for the composition. 
The low MARs of A0 therefore cannot be attributed primarily to failure to transcribe or understand the spoken instruction.

\noindent\textbf{Comparison among A1 strategies.}
One-shot prompting is the strongest A1 adaptation for both standalone tasks, reaching 19\% on \textsc{Hipi} and 17\% on \textsc{Yingchen}. Human-mimic prompting and mechanism-aware prompting are competitive on \textsc{Yingchen} at 16\% and 15\%. Cross-model aggregation is the least effective, achieving only 4\% and 10\%  on the two tasks. A concrete input--response demonstration is therefore more useful to the attacker than a verbal explanation or output aggregation, although it still leaves most challenges rejected. 

\subsection{Human Usability}
\label{sec:rq4-usability}

\subsubsection{Study Design}
We recruit 50 adult native English speakers from English-speaking countries through CloudResearch Connect~\cite{cloudresearch2026connect}. We use the platform's participation-history controls to exclude individuals who had participated in the pilot study.

\noindent \textbf{Questionnaire Design.} 
Each participant completes one questionnaire containing one challenge from each of the 11 audio CAPTCHA schemes, including \textsc{Hipi}, \textsc{Yingchen}, and nine baselines, presented in randomized order. All challenge instances are distinct across participants, yielding 50 distinct instances per CAPTCHA type.

\noindent \textbf{Quality Control.} 
Each questionnaire also includes three randomly inserted attention checks requiring participants to enter an instructed character (``a'', ``b'', or ``c''). Participants failing any attention check are excluded. 
Participants must listen to each challenge audio in full at least once and enter an answer before proceeding, but may replay the audio without restriction.
Participants passing the attention check receive one dollar upon completion. 

\noindent \textbf{Evaluation Metrics.}
We measure \emph{effectiveness} by HAR, \emph{efficiency} by completion time from the initial playback action to final submission and replay rate, and \emph{satisfaction}. Replay rate is the proportion of challenges for which the audio is played more than once. Participants also rate their satisfaction with each of the 11 CAPTCHA schemes on a five-point Likert scale~\cite{likert1932technique}: \emph{very dissatisfied} (1), \emph{dissatisfied} (2), \emph{neither dissatisfied nor satisfied} (3), \emph{satisfied} (4), and \emph{very satisfied} (5).

{To evaluate the composite CAPTCHA, the \textsc{Hipi} and \textsc{Yingchen} trials completed by each participant form one within-participant pair, yielding 50 pairs. A pair is accepted only if both component responses are accepted; its completion time is the sum of the two component times, and it is counted as replayed if either component is replayed. Further procedural details are provided in Appendix~\ref{app:human-questionnaire}.}

\subsubsection{Study Results}
{All 50 participants passed all three attention checks
and were retained for analysis. The results are shown in
\tablename~\ref{tab:human-evaluation-full}.
}

\noindent\textbf{Effectiveness.}
\textsc{Hipi} and \textsc{Yingchen} achieve HARs of 78.0\% and 90.0\%, respectively. Compared with the nine baselines, \textsc{Hipi} outperforms five, while \textsc{Yingchen} outperforms six and matches Math, demonstrating competitive human acceptance rates. Their composition achieves a 76.0\% HAR, only 2 percentage points below \textsc{Hipi} and 14 percentage points below \textsc{Yingchen}, while outperforming five baselines. 
Recall that the composition reduces the maximum observed MAR from 19\% and 17\% for the standalone tasks to just 3\%. These results demonstrate that conjunctive verification substantially strengthens machine resistance while preserving high human acceptance.

\noindent\textbf{Efficiency.}
The median completion times of \textsc{Hipi} and \textsc{Yingchen} are 51.1~s and 28.6~s, respectively. The longer completion time of \textsc{Hipi} likely reflects both its greater task difficulty and longer challenge audio duration. Notably, \textsc{Yingchen} achieves the lowest replay rate (12.0\%) among all evaluated CAPTCHAs, while \textsc{Hipi} achieves 22.0\%, outperforming seven of the nine baselines. Their composition has a median completion time of 81.7~s and a replay rate of 28.0\%, still lower than seven baselines. These results demonstrate that the proposed CAPTCHAs require relatively few replays, even when both tasks are combined, with \textsc{Yingchen} offering particularly efficient interaction.

\noindent\textbf{Satisfaction.}
\textsc{Hipi} and \textsc{Yingchen} receive satisfaction ratings of 2.68/5 and 3.10/5, respectively, outperforming five and six of the nine baselines, respectively. Both also outperform all four 
advanced baselines (adversarial audio- or perception-mismatch-based): aaeCAPTCHA, Yeehaw, Meutzner, and Volcano. 
The higher satisfaction of \textsc{Yingchen} is consistent with its higher HAR, shorter completion time, and lower replay rate.

Overall, the results demonstrate competitive usability across effectiveness, efficiency, and satisfaction. The proposed designs offer complementary deployment choices, with \textsc{Yingchen} providing lighter interaction and the composition delivering the strongest machine resistance.

\begin{table}[t]
\centering
\caption{Human usability results.}
\label{tab:human-evaluation-full}
\small
\setlength{\tabcolsep}{4pt}
\renewcommand{\arraystretch}{1.00}

\resizebox{\columnwidth}{!}{%
\begin{tabular}{@{}llcccc@{}}
    \toprule
    \multicolumn{2}{c}{Audio CAPTCHAs}
    & HAR
    & Median time
    & Replay rate
    & Satisfaction \\
    &
    & (\%)
    & (s)
    & (\%)
    & (1--5) \\
    \midrule

    \multirow{3}{*}{Ours}
    & \textsc{Hipi}
    & 78.0
    & 51.1
    & 22.0
    & 2.68 \\
    & \textsc{Yingchen}
    & 90.0
    & 28.6
    & 12.0
    & 3.10 \\
    & Composite
    & 76.0
    & 81.7
    & 28.0
    & -- \\

    \midrule

    \multirow{9}{*}{Baselines}
    & Control
    & 18.0
    & 24.9
    & 48.0
    & 1.96 \\
    & Math
    & 90.0
    & 16.3
    & 18.0
    & 3.34 \\
    & Character
    & 84.0
    & 21.3
    & 34.0
    & 3.06 \\
    & Pauses
    & 94.0
    & 22.7
    & 18.0
    & 4.10 \\
    & Categories
    & 96.0
    & 25.9
    & 32.0
    & 3.38 \\
    & aaeCAPTCHA
    & 12.0
    & 36.6
    & 68.0
    & 1.78 \\
    & Yeehaw
    & 24.0
    & 28.5
    & 74.0
    & 1.66 \\
    & Meutzner
    & 38.0
    & 41.6
    & 50.0
    & 1.50 \\
    & Volcano
    & 6.0
    & 31.3
    & 66.0
    & 1.70 \\

    \bottomrule
\end{tabular}%
}
\end{table}

\subsection{Physical Ground Truth in Rejected LALM Responses}
\label{sec:physical-answer-analysis}

The MAR evaluation in \Cref{sec:rq3-security} measures agreement with human-reference answers. Among rejected challenges, however, LALM responses may agree with physical ground truth or differ from both reference answers. This subsection examines the extent to which rejected responses reflect physical ground truth, characterizing physical-recovery tendencies when the CAPTCHA successfully resists an attack.

\subsubsection{Task-Specific Physical Agreement Metrics}
\label{sec:challenge-physical-accuracy}

For each task and solver--strategy configuration, $N$ denotes the number of rejected challenges.

\noindent \textbf{Label matching rate (LMR) for \textsc{Hipi}.}
For the classification-based \textsc{Hipi}, a challenge is rejected when at least one predicted stimulus label differs from its human-reference label. A nonmatching label may either agree with the physical label or differ from both reference labels. LMR quantifies physical-label agreement across all stimuli in the rejected challenges:
\begin{equation}
\mathrm{LMR}_{S,a}
=
\frac{1}{NK}
\sum_{i=1}^{N}
\sum_{j=1}^{K}
\mathbb{I}
\left[
y_{S,a}^{(j)}(x_H^i)=y^\star(s_i^j)
\right].
\end{equation}
Here, $K$ is the number of stimuli per challenge, $s_i^j$ is the $j$-th stimulus of challenge $x_H^i$, and $y_{S,a}^{(j)}(x_H^i)$ is its parsed solver label.

\begin{table*}[t]
\centering
\caption{Physical tendencies among rejected challenges (\%).
Each entry reports \textsc{Hipi} LMR / \textsc{Yingchen} OCR.}
\label{tab:physical-answer-analysis}

\footnotesize
\setlength{\tabcolsep}{4pt}
\renewcommand{\arraystretch}{1.12}

\resizebox{\textwidth}{!}{%
\begin{tabular}{@{}lccccccc@{}}
\toprule
Solver & A0-Audio & A0-Text & A1-CR & A1-Mec
       & A1-Mimic & A1-One-shot & A1-Vote \\
\midrule

gpt-realtime-2
& 22.94 / 29.55 & 23.92 / 37.78 & 19.59 / 43.48
& 19.33 / 41.11 & 20.31 / 44.19 & 17.78 / 68.24
& 20.21 / 27.27 \\

gpt-audio-1.5
& 0.00 / 0.00 & 0.00 / 0.00 & 4.25 / 0.00
& 0.00 / 0.00 & 8.50 / 14.77 & 11.73 / 34.52
& 0.00 / 0.00 \\

qwen3.5-omni-plus
& 11.70 / 47.78 & 12.77 / 53.93 & 12.90 / 47.42
& 11.65 / 59.34 & 15.73 / 49.44 & 17.59 / 100.00
& 13.06 / 59.09 \\

qwen3.5-omni-flash 
& 38.38 / 29.55 & 39.25 / 28.74 & 44.25 / 39.33
& 43.25 / 28.24 & 29.00 / 28.57 & 32.61 / 89.66
& 39.39 / 30.23 \\

kimi-audio-7b
& 28.00 / 0.00 & 14.03 / 7.78 & 12.24 / 6.82
& 12.37 / 6.74 & 15.15 / 15.73 & 17.25 / 18.07
& 18.43 / 9.20 \\

nemotron-3-nano-omni
& 10.75 / 5.43 & 20.75 / 10.64 & 23.00 / 1.06
& 20.96 / 5.32 & 26.25 / 0.00 & 18.18 / 88.64
& 25.50 / 5.56 \\

qwen2-audio-7b
& 0.00 / 0.00 & 0.00 / 0.00 & 0.00 / 10.00
& 0.00 / 6.38 & 13.50 / 1.03 & 14.50 / 79.78
& 1.00 / 7.69 \\

\bottomrule
\end{tabular}%
}
\end{table*}

\noindent \textbf{Over-counting rate (OCR) for \textsc{Yingchen}.}
For the counting-based \textsc{Yingchen}, a challenge is rejected when the predicted count differs from the human-reference count. Rejected responses therefore fall into two cases: counts below or above the human reference. Because the physical count exceeds the human-reference count, the latter is consistent with a counting tendency toward physical ground truth. OCR measures the proportion of rejected responses in this second case:
\begin{equation}
\mathrm{OCR}_{S,a}
=
\frac{1}{N}
\sum_{i=1}^{N}
\mathbb{I}
\left[
y_{S,a}(x_Y^i)>y_H^\dagger(x_Y^i)
\right].
\end{equation}

\subsubsection{Results}
\label{sec:partial-physical-responses}
The results are shown in Table~\ref{tab:physical-answer-analysis}. 

\noindent \textbf{\textsc{Hipi}.}
Five out of seven models exhibit nonzero LMRs across all attack strategies, with values ranging from 10.75\% to 44.25\%. 
For instance, qwen3.5-omni-flash achieves LMRs of 38.38\%, 44.25\%, and 43.25\% under A0-Audio, A1-CR, and A1-Mec, respectively. 
These results indicate that LALMs exhibit a tendency to recover physical stimulus labels. 

\noindent \textbf{\textsc{Yingchen}.}
Three models exhibit nonzero OCRs across all attack strategies, ranging from 27.27\% to 100\%. 
Under A1-One-shot, qwen3.5-omni-plus over-counts all rejected challenges, yielding an OCR of 100\%; its OCRs under the other strategies range from 47.42\% to 59.34\%. 
Similarly, 
qwen3.5-omni-flash, nemotron-3-nano-omni, and qwen2-audio-7b reach 89.66\%, 88.64\%, and 79.78\% under A1-One-shot, respectively. These observations indicate a pronounced tendency to report counts above the human reference, in the direction of the larger physical count. 

Zero LMRs occur for gpt-audio-1.5 and qwen2-audio-7b under several strategies, while zero OCRs also occur for these models, kimi-audio-7b, and nemotron-3-nano-omni. 
Such configurations may arise when LALMs fail to resolve pitch direction in \textsc{Hipi} or, in \textsc{Yingchen}, miss low-salience sounds or merge separate segments. 
Nevertheless, rejection depends on disagreement with the human-reference answer. Responses differing from both references therefore remain rejected, providing a fallback rejection path when physical-ground-truth recovery is limited.

%% file: discussion_new.tex
\section{Discussion}
\label{sec:discussion}

\noindent\textbf{Individual Perceptual Variability.}
Reverse-capability verification uses the modal human response as its reference and therefore represents the dominant population-level response rather than every listener's perception. Individual differences in hearing thresholds, temporal resolution, and the interpretation of ambiguous sounds may cause some users to recover the physical ground truth or report a different perceptual outcome~\cite{moore2012psychologyhearing,phillips1999gap,malek2018pitchclass}. 
Such responses would not match the human-reference answer and would therefore be rejected. Consequently, the observed HARs may not generalize to populations whose auditory characteristics are underrepresented.

\noindent\textbf{Broader Reverse-Capability Instantiations.}
Our implementation draws on psychoacoustics, but the reverse-capability framework is not limited to psychoacoustic mechanisms or the audio modality. Within audio CAPTCHA design, future work could investigate whether higher-level linguistic, semantic, or cognitive processing of auditory content can produce the required stable response divergence. The same principle could be extended beyond audio to visual and multimodal settings, including visual illusions, temporal-order perception, and cross-modal integration. Compatible mechanisms within and across modalities could be combined under a joint verifier, potentially increasing challenge diversity and reducing dependence on any single mechanism. 

\noindent\textbf{Attacker-Model Scope and Stronger Adaptive Attacks.}
Our security evaluation is limited to general-purpose LALMs used as black-box solvers with inference-time adaptation, excluding learned or optimization-based adaptation using labeled CAPTCHA data or white-box model access. 
Such adaptation could more directly capture the mapping from acoustic cues to verifier-accepted human responses. 
Future work should therefore evaluate mechanism-specific solver fine-tuning and automated optimization of prompts, adversarial suffixes, or audio transformations through gradients, surrogate models, or black-box search to increase the probability of producing $y_H^\dagger(x)$.

%% file: conclusion_new.tex
\section{Conclusion}
\label{sec:conclusion}

This work introduced reverse-capability verification, which exploits stable perceptual limitations so that the verifier accepts a human-reference answer differing from the physical ground truth machines tend to recover. We instantiated the framework through psychoacoustics as \textsc{Hipi}, \textsc{Yingchen}, and their conjunctive composition. Our security evaluation and human usability study show lower machine acceptance than prior schemes while preserving high human acceptance, with the composition providing the strongest security. Thus, CAPTCHA security need not rely on machines being worse than humans at recovering the physical ground truth, making reverse-capability verification a promising alternative to conventional CAPTCHA design.

%% file: appendices.tex
\appendices

\section{More Details of the Pilot Study}
\label{app:pilot_materials}

\subsection{Stimulus-Class Generation Parameters}
\label{app:formative-materials}

Table~\ref{tab:stimulus-generation-parameters} summarizes the generation parameters for the stimulus classes used in the pilot study and CAPTCHA construction. All stimuli were mono, 48-kHz, 24-bit PCM. 

\begin{table*}[!tp]
    \centering
    \caption{Generation parameters of stimulus classes.}
    \label{tab:stimulus-generation-parameters}
    \fontsize{7}{7.5}\selectfont
    \setlength{\tabcolsep}{3.0pt}
    \renewcommand{\arraystretch}{0.90}

    \begin{threeparttable}
    \begin{tabularx}{\textwidth}{
        @{}
        p{0.22\textwidth}
        p{0.12\textwidth}
        p{0.20\textwidth}
        X
        @{}
    }
        \toprule
        Stimulus class & Use & Parameter & Value / sampling range \\
        \midrule

        \multirow{10}{0.22\textwidth}{Shepard--Risset glissando}
        & \multirow{10}{0.12\textwidth}{Pilot; \textsc{Hipi}}
        & Direction & Condition-controlled: up or down \\
        & & Duration & 7~s \\
        & & Octave components & 10 \\
        & & Spectral range & 19.6--20,070.4~Hz \\
        & & Envelope center & \(U(200,250)\)~Hz \\
        & & Envelope width & \(U(1.25,1.60)\) octaves \\
        & & Starting reference & \(U(180,320)\)~Hz \\
        & & Component phase & Independent \(U(0,2\pi)\) per component \\
        & & Outer ramp & 25~ms \\
        & & RMS & 0.105 \\
        \midrule

        \multirow{8}{0.22\textwidth}{Glide-disambiguated Shepard tritone}
        & \multirow{8}{0.12\textwidth}{Pilot only}
        & Direction & Condition-controlled: up or down \\
        & & Endpoints & Uniformly sampled transposition of a tritone pair \\
        & & Anchor duration & Shared \(U(1.8,2.2)\)~s \\
        & & Glide duration & \(U(2.7,3.3)\)~s \\
        & & Octave components & 10 \\
        & & Component phase & Independent \(U(0,2\pi)\) per component \\
        & & Outer ramp & 20~ms \\
        & & RMS & 0.105 \\
        \midrule

        \multirow{9}{0.22\textwidth}{Missing fundamental}
        & \multirow{9}{0.12\textwidth}{Pilot only}
        & Fundamental & \(U(220,280)\)~Hz \\
        & & Highest harmonic \(H\) & \(U_d(\{5,6,7\})\) \\
        & & Reference harmonics & 1--\(H\) \\
        & & Candidate harmonics & 2--\(H\) \\
        & & Tone duration & Shared \(U(0.9,1.1)\)~s \\
        & & Interstimulus gap & \(U(0.4,0.6)\)~s \\
        & & Partial phase & Independent \(U(0,2\pi)\) per partial \\
        & & Outer ramp & 20~ms \\
        & & Interval RMS & 0.08 \\
        \midrule

        \multirow{8}{0.22\textwidth}{Zwicker tone}
        & \multirow{8}{0.12\textwidth}{Pilot only}
        & Noise duration & \(U(4.5,5.5)\)~s \\
        & & Observation interval & \(U(5.0,6.0)\)~s \\
        & & Noise band & 30--16,000~Hz \\
        & & Notch center & Log-uniform sampling over 3--4~kHz \\
        & & Notch width & \(U(0.50,0.70)\) octaves \\
        & & Outer ramp & 20~ms \\
        & & Noise RMS & 0.08 \\
        & & Noise realization & Independently seeded Gaussian noise \\
        \midrule

        \multirow{9}{0.22\textwidth}{Small-span high-frequency glide}
        & \multirow{9}{0.12\textwidth}{Pilot; \textsc{Hipi}}
        & Direction & Condition-controlled: up or down \\
        & & Duration & 7~s \\
        & & Initial frequency & \(U(3,5)\)~kHz \\
        & & Displacement & Direction sign \(\times U(20,28)\) cents \\
        & & Frequency ratios & 1:2:3:4 \\
        & & Amplitude ratios & 1:0.35:0.16:0.08, independently perturbed by \(U(0.9,1.1)\) and renormalized \\
        & & Component phase & Independent \(U(0,2\pi)\) per component \\
        & & Outer ramp & 25~ms \\
        & & RMS & 0.105 \\
        \midrule

        \multirow{10}{0.22\textwidth}{Moving spectral-ridge noise}
        & \multirow{10}{0.12\textwidth}{Pilot; \textsc{Hipi}}
        & Direction & Condition-controlled: up or down \\
        & & Duration & 7~s \\
        & & Frame/hop & 2,048/512 samples \\
        & & Passband & 120--8,000~Hz \\
        & & Ridge centers & \(U(450,650)\) and \(U(2400,2800)\)~Hz; order set by direction \\
        & & Ridge width & \(U(0.55,0.75)\) octaves \\
        & & Contrast & \(U(16,20)\)~dB \\
        & & Jitter SD/window & \(U(0.22,0.34)\) octaves / \(U_d(\{25,31,37\})\) frames \\
        & & Ramp / RMS & 25~ms / 0.105 \\
        & & Noise & Independently seeded Gaussian \\
        \midrule

        \multirow{6}{0.22\textwidth}{Audible control}
        & \multirow{6}{0.12\textwidth}{\textsc{Yingchen} only}
        & Duration & 1~s \\
        & & Profile & \(U_d(\{\text{medium},\text{narrow}\})\) \\
        & & Center/bandwidth & Medium: \(U(700,800)/U(80,120)\)~Hz; narrow: \(U(475,525)/U(20,30)\)~Hz \\
        & & Outer ramp & 20~ms \\
        & & RMS & 0.08 \\
        & & Carrier realization & Independently seeded Gaussian noise \\
        \midrule

        \multirow{5}{0.22\textwidth}{Very-low-frequency tone}
        & \multirow{5}{0.12\textwidth}{Pilot; \textsc{Yingchen}}
        & Duration & 1~s \\
        & & Frequency & \(U_d(23{:}0.01{:}24)\)~Hz \\
        & & Relative level & \(U_d(-10{:}0.1{:}-5)\)~dB \\
        & & Initial phase & \(U(0,2\pi)\) \\
        & & Outer ramp & 100~ms \\
        \midrule

        \multirow{6}{0.22\textwidth}{Strongly attenuated sound}
        & \multirow{6}{0.12\textwidth}{Pilot; \textsc{Yingchen}}
        & Duration & 1~s \\
        & & Profile & \(U_d(\{\text{medium},\text{narrow}\})\) \\
        & & Center/bandwidth & Medium: \(U(700,800)/U(80,120)\)~Hz; narrow: \(U(475,525)/U(20,30)\)~Hz \\
        & & Reference RMS & 0.08 \\
        & & Relative level & \(U_d(-68{:}0.1{:}-65)\)~dB \\
        & & Carrier realization & Independently seeded Gaussian noise \\
        \midrule

        \multirow{8}{0.22\textwidth}{Very-short silent gap}
        & \multirow{8}{0.12\textwidth}{Pilot; \textsc{Yingchen}}
        & Duration & 1~s \\
        & & Profile & \(U_d(\{\text{medium},\text{narrow}\})\) \\
        & & Center/bandwidth & Medium: \(U(700,800)/U(80,120)\)~Hz; narrow: \(U(475,525)/U(20,30)\)~Hz \\
        & & RMS & 0.08 \\
        & & Silent-gap length & \(U_d(\{5,\ldots,120\})\) samples (0.104--2.5~ms) \\
        & & Gap ramp & 12 samples (0.25~ms) per side \\
        & & Gap center & \(U(0.35,0.65)\)~s \\
        & & Carrier realization & Independently seeded Gaussian noise \\
        \bottomrule
    \end{tabularx}

    \begin{tablenotes}
        \scriptsize
        \item Note: We write \(U(a,b)\) for continuous uniform sampling and \(U_d(S)\) for uniform sampling from the set \(S\). We write \(a{:}\Delta{:}b\) for the discrete grid \(\{a,a+\Delta,\ldots,b\}\). Unless otherwise stated, sampled quantities were independently redrawn for each stimulus instance, whereas bare values denote fixed parameters. Reported dB values denote digital level ratios.
    \end{tablenotes}
    \end{threeparttable}
\end{table*}

\subsection{Pilot Questions}
\label{app:pilot-questions}

\begin{figure*}[t]
    \centering

    \begin{minipage}[t]{0.31\textwidth}
        \centering
        \includegraphics[width=\linewidth]{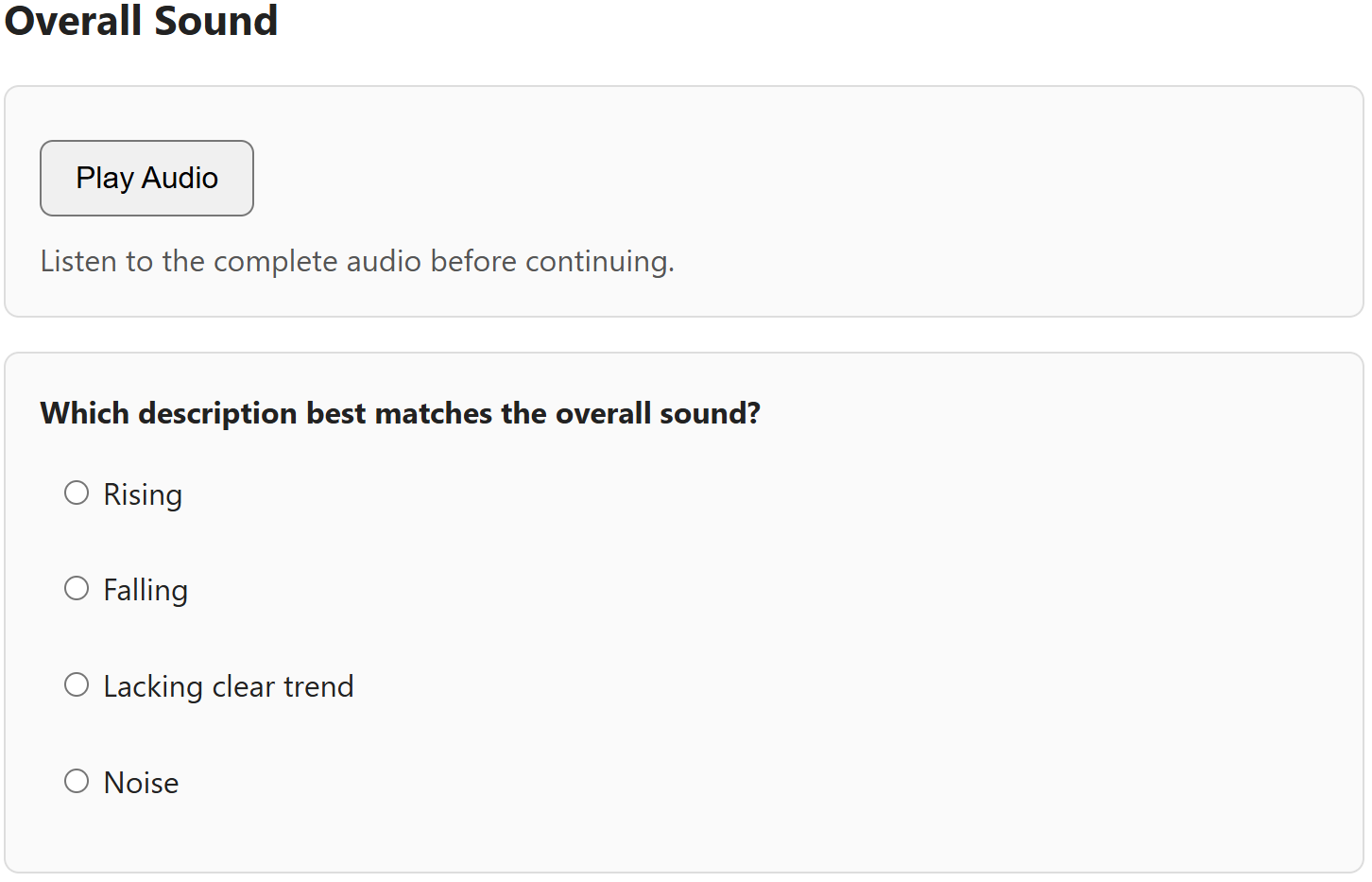}
        \small (a) Overall sound
    \end{minipage}
    \hfill
    \begin{minipage}[t]{0.31\textwidth}
        \centering
        \includegraphics[width=\linewidth]{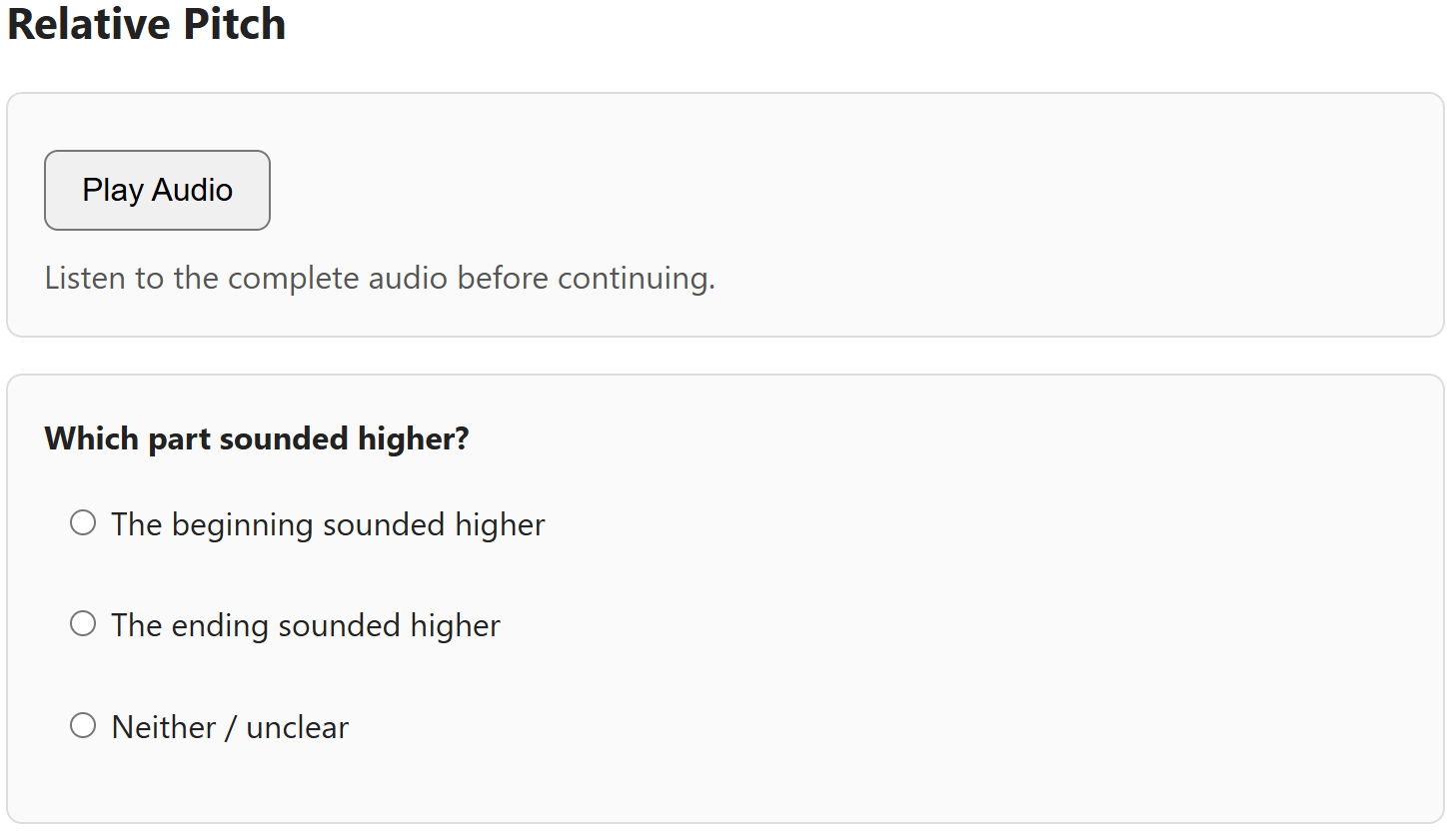}
        \small (b) Relative pitch
    \end{minipage}
    \hfill
    \begin{minipage}[t]{0.31\textwidth}
        \centering
        \includegraphics[width=\linewidth]{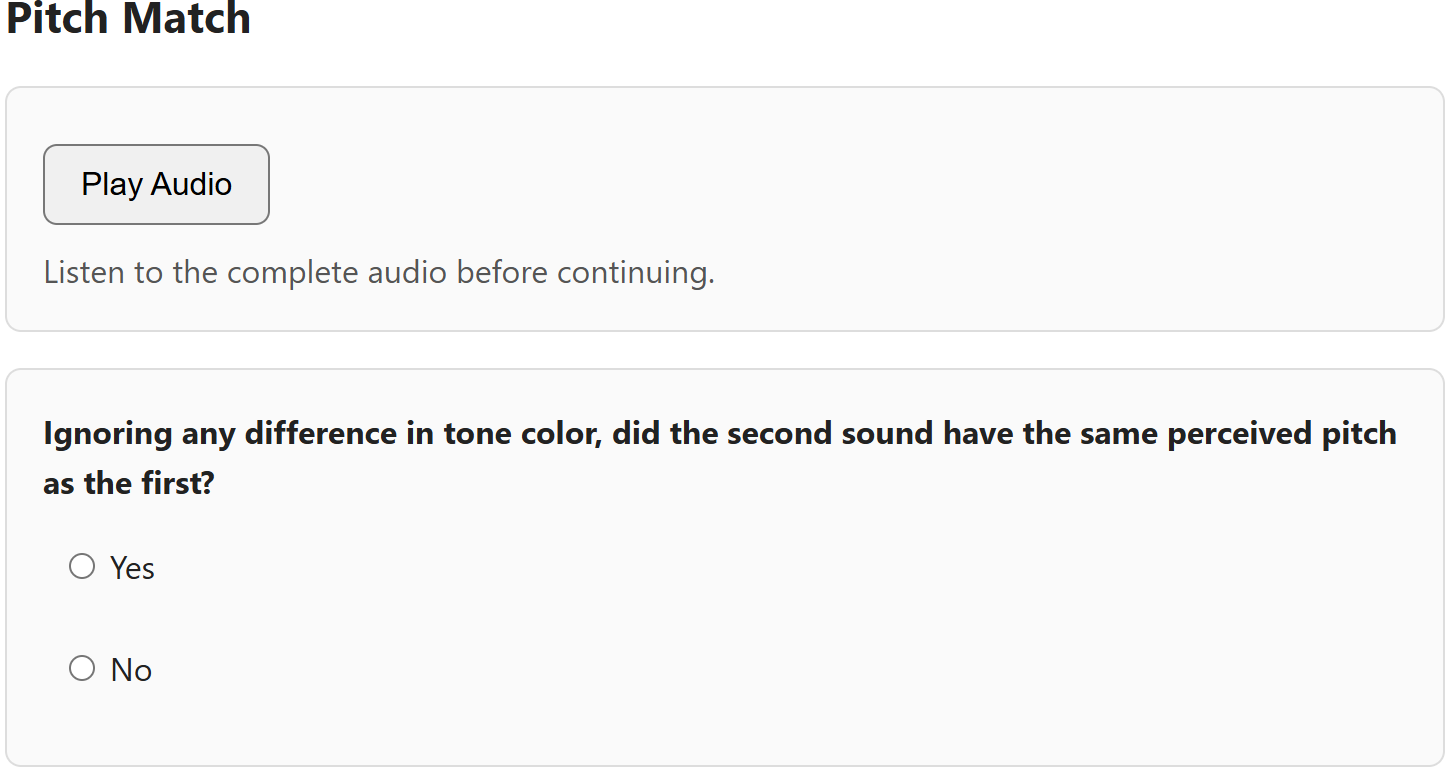}
        \small (c) Pitch match
    \end{minipage}

    \vspace{0.8em}

    \begin{minipage}[t]{0.31\textwidth}
        \centering
        \includegraphics[width=\linewidth]{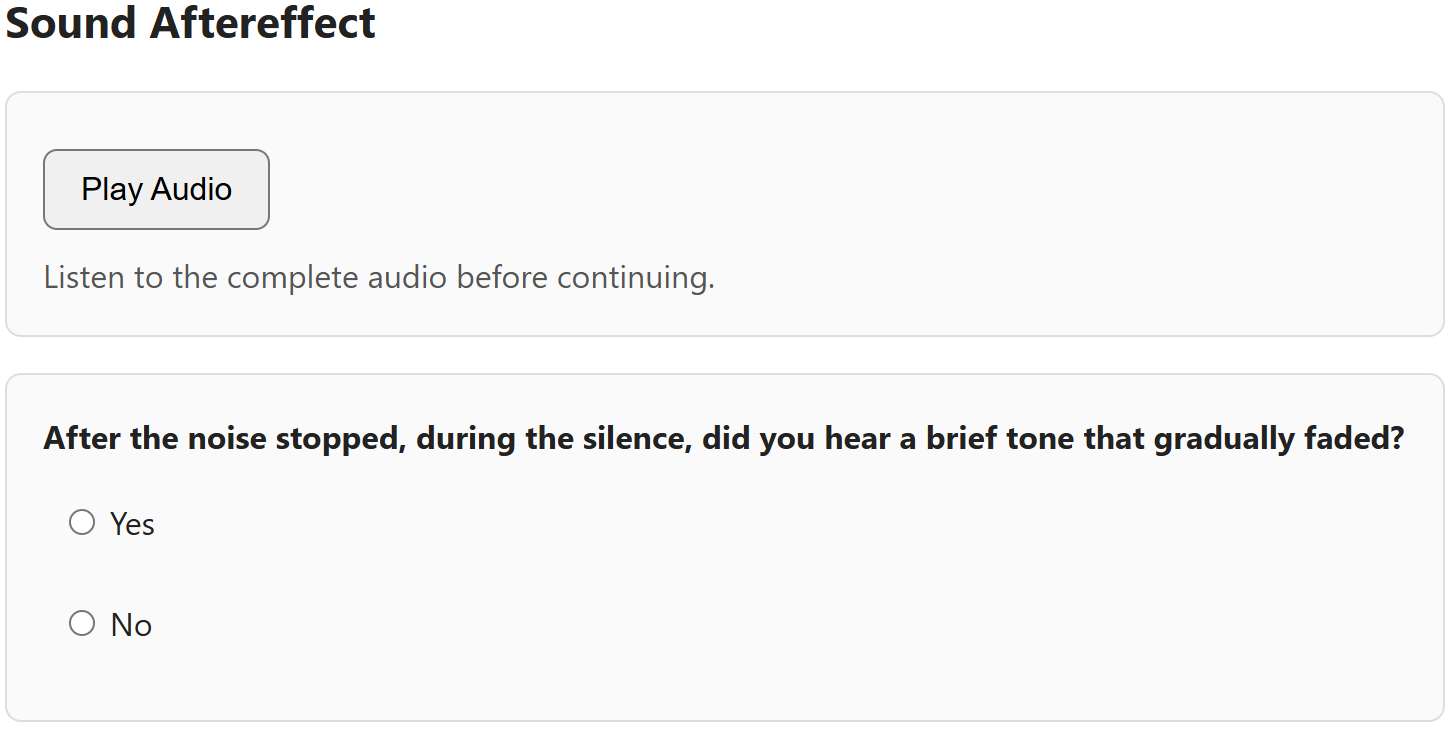}
        \small (d) Sound aftereffect
    \end{minipage}
    \hspace{0.04\textwidth}
    \begin{minipage}[t]{0.31\textwidth}
        \centering
        \includegraphics[width=\linewidth]{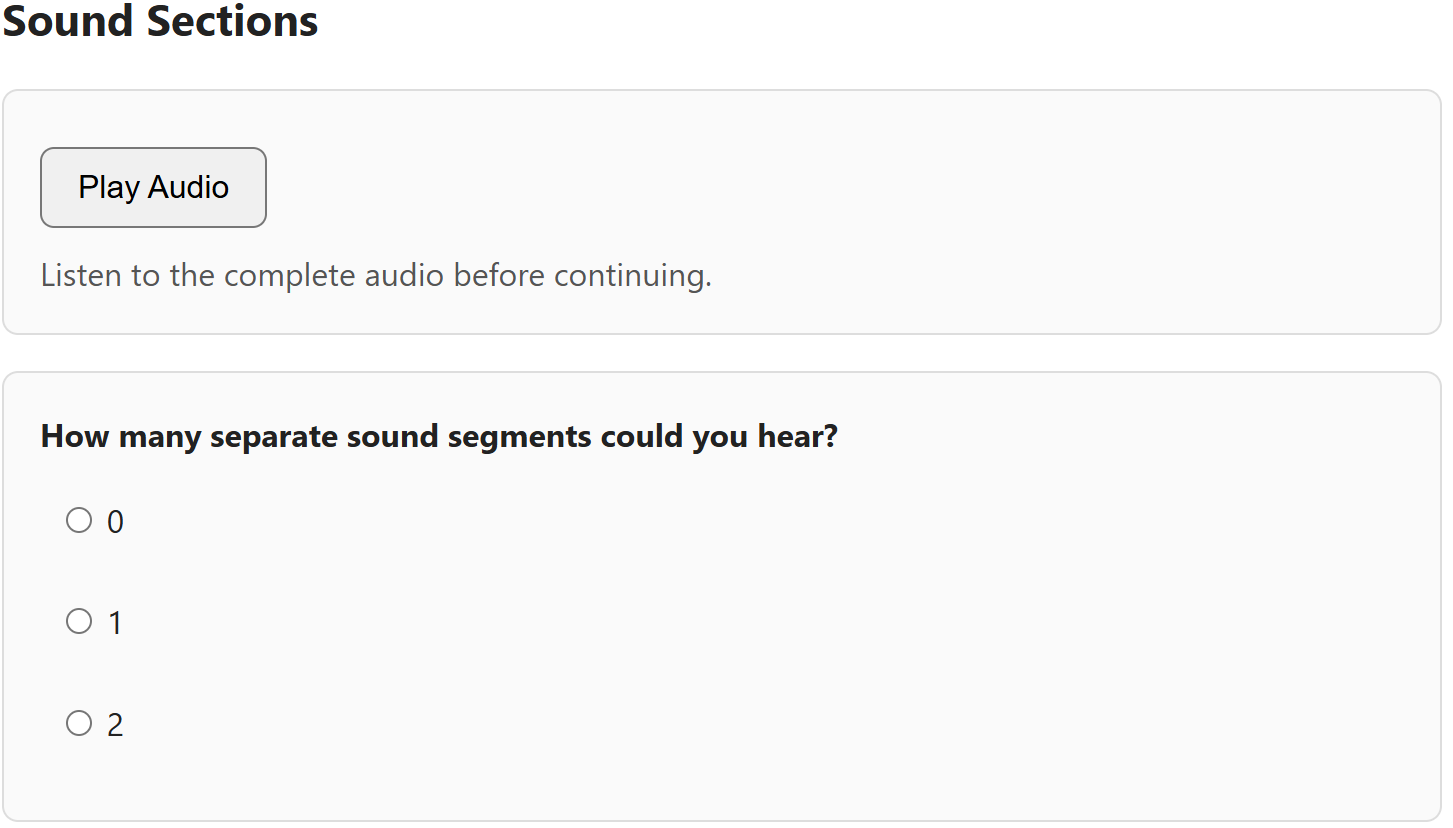}
        \small (e) Sound sections
    \end{minipage}

    \caption{Participant-facing questions used in the pilot study.}
    \label{fig:pilot-questions}
\end{figure*}

The pilot study used stimulus-specific questions matched to the perceptual judgment required by each candidate class. Figure~\ref{fig:pilot-questions} shows the participant-facing questions and response options used in the study. The overall-sound question was used for the Shepard--Risset glissando, small-span high-frequency glide, and moving spectral-ridge noise conditions; the relative-pitch question for the glide-disambiguated Shepard tritone; the pitch-match question for the missing-fundamental condition; the sound-aftereffect question for the Zwicker-tone condition; and the sound-sections question for the very-low-frequency tone, strongly attenuated sound, and very-short silent-gap conditions.

\section{CAPTCHA Implementation Details}
\label{app:captcha_implementation}

\subsection{Response Parsing}
\label{app:parsing}

For \textsc{Hipi}, permitted separators are whitespace, commas, semicolons, colons, slashes, vertical bars, periods, and hyphens. Labels are not extracted from prose, JSON objects, explanations, or longer strings.

For \textsc{Yingchen}, signed values, decimals, number words, explanations, responses containing additional numbers, and otherwise ambiguous outputs are rejected.

\subsection{CAPTCHA Assembly}
\label{app:audio-assembly}

Spoken instructions were synthesized using the Microsoft David Desktop (en-US) text-to-speech voice~\cite{microsoftVoiceInformation} at SAPI rate \(-1\) and normalized to an RMS level of 0.075. For \textsc{Hipi}, every silent separator, including the interval between the instruction and the first stimulus and those between adjacent stimuli, was 0.45~s. For \textsc{Yingchen}, the interval between the instruction and the first stimulus was 0.45~s, while each inter-stimulus silent separator was independently sampled from 0.20 to 0.30~s.

For \textsc{Hipi}, the instruction was:

\begin{quote}
\textit{The test sequence contains four segments. Assign one label to each segment: R for rising, F for falling, L for no clear overall trend, or N for noise. Return four letters in order.}
\end{quote}

For \textsc{Yingchen}, the instruction was:

\begin{quote}
\textit{A test sequence follows. Determine the number of segments separated by silent intervals. Answer with one integer.}
\end{quote}

\section{Machine-Evaluation Materials}
\label{app:machine_materials}

\subsection{Attack Prompts and Demonstrations}
\label{app:attack-prompts}

We provide the exact textual inputs used in the evaluated prompt-based attack conditions. A0-Audio uses no additional textual input, while A1-Vote and A1-Ensemble use the A0-Text input without additional prompt text.

\noindent\textbf{A0-Text.}
A0-Text used the verbatim instruction transcripts reported in Appendix~\ref{app:audio-assembly}, with no additional text.

\noindent\textbf{A1-CR.}
For \textsc{Hipi}, the prompt was:

\begin{promptbox}
\textbf{Prompt:} The test sequence contains four segments. Assign one label to each segment: R for rising, F for falling, L for no clear overall trend, or N for noise. Return four letters in order. Carefully inspect each segment and perform a final self-check before answering. Return exactly four letters and nothing else.
\end{promptbox}

For \textsc{Yingchen}, the prompt was:

\begin{promptbox}
\textbf{Prompt:} A test sequence follows. Determine the number of segments separated by silent intervals. Answer with one integer. Carefully inspect the entire sequence and perform a final self-check before answering. Return only one integer.
\end{promptbox}

\noindent\textbf{A1-Mec.}
For \textsc{Hipi}, the prompt was:

\begin{promptbox}
\textbf{Prompt:} The test sequence contains four segments. Assign one label to each segment: R for rising, F for falling, L for no clear overall trend, or N for noise. Return four letters in order. The CAPTCHA is constructed from three types of psychoacoustic stimuli: Shepard--Risset glissandi composed of octave-spaced components whose amplitudes are shaped as their frequencies move; high-band frequency glides containing only a slight monotonic frequency displacement; and moving spectral-ridge noise in which a broad region of spectral energy moves upward or downward over time. Use this mechanism information to analyze each segment. Return exactly four letters and nothing else.
\end{promptbox}

For \textsc{Yingchen}, the prompt was:

\begin{promptbox}
\textbf{Prompt:} A test sequence follows. Determine the number of segments separated by silent intervals. Answer with one integer. The CAPTCHA is constructed from four types of acoustic stimuli: clearly audible sounds; very-low-frequency sounds; very-low-level sounds; and sounds containing very short silent gaps. The latter three manipulate frequency, signal level, and temporal structure, respectively. Use this mechanism information to analyze the sequence. Return only one integer.
\end{promptbox}

\noindent\textbf{A1-Mimic.}
For \textsc{Hipi}, the prompt was:

\begin{promptbox}
\textbf{Prompt:} The test sequence contains four segments. Assign one label to each segment: R for rising, F for falling, L for no clear overall trend, or N for noise. Return four letters in order. Respond as an ordinary untrained listener using normal playback would after listening once. Predict the four labels that such a listener would most likely submit, rather than labels obtained through detailed technical analysis. Return exactly four letters and nothing else.
\end{promptbox}

For \textsc{Yingchen}, the prompt was:

\begin{promptbox}
\textbf{Prompt:} A test sequence follows. Determine the number of segments separated by silent intervals. Answer with one integer. Respond as an ordinary untrained listener using normal playback would after listening once. Predict the count that such a listener would most likely submit, rather than every boundary detectable through detailed forensic analysis. Return only one integer.
\end{promptbox}

\noindent\textbf{A1-One-shot.}
For each CAPTCHA, the same fixed demonstration was used across all solvers and test challenges. The demonstration and test challenge were presented in the following three-message sequence.

For \textsc{Hipi}, the message sequence was:

\begin{promptbox}
\textbf{User:} The test sequence contains four segments. Assign one label to each segment: R for rising, F for falling, L for no clear overall trend, or N for noise. Return four letters in order.

\textbf{Demonstration 1.}

\texttt{[demonstration audio]}

\medskip
\textbf{Assistant:} \texttt{RFLN}

\medskip
\textbf{User:} Test item. Return only the answer in the required format.

\texttt{[query audio]}
\end{promptbox}

For \textsc{Yingchen}, the message sequence was:

\begin{promptbox}
\textbf{User:} A test sequence follows. Determine the number of segments separated by silent intervals. Answer with one integer.

\textbf{Demonstration 1.}

\texttt{[demonstration audio]}

\medskip
\textbf{Assistant:} \texttt{8}

\medskip
\textbf{User:} Test item. Return only the answer in the required format.

\texttt{[query audio]}
\end{promptbox}

\subsection{Request and Failure Handling}
\label{app:failure-handling}

Transport- or provider-level failures were retried until a response was returned, with a 180-s timeout applied to each request. Once an application-level response was returned, the request was not retried: empty responses, refusals, and syntactically invalid or unparseable outputs were mapped to \(\bot\), while valid but incorrect answers were rejected. Any completed model invocation that returned an application-level response counted toward the query budget, including responses subsequently mapped to \(\bot\); transport- or provider-level attempts that returned no response did not.

For A1-Vote and A1-Ensemble, each model output was first processed by the corresponding response parser. Outputs mapped to \(\bot\) did not participate in aggregation. Aggregation was then performed over the remaining parsed responses. If multiple participating responses were tied for the highest frequency, one of the tied responses was selected at random. If no valid parsed responses remained, the aggregated response was mapped to $\bot$.

\begin{figure}[t]
    \centering
    \includegraphics[width=\columnwidth]{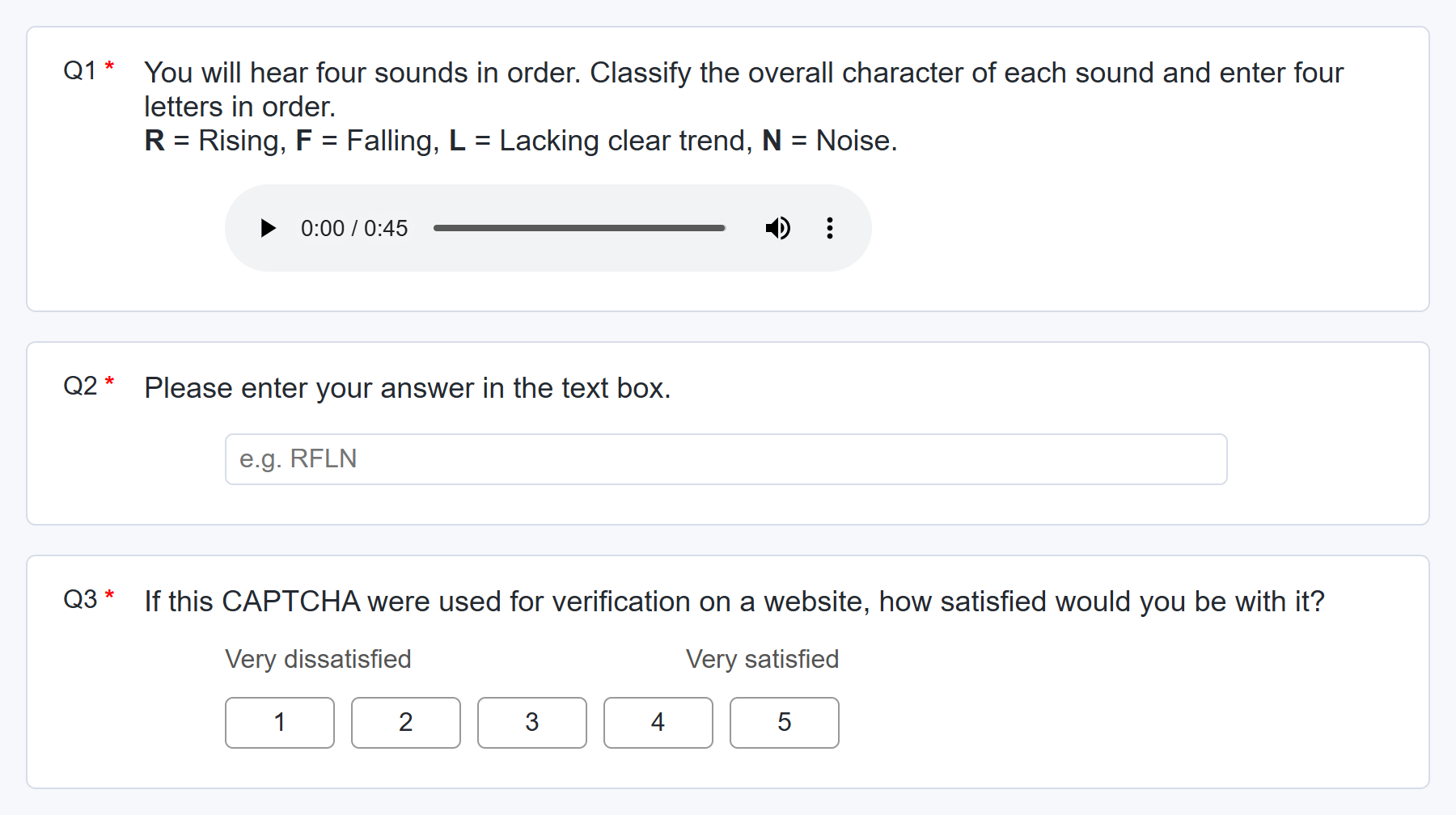}
    \caption{Participant interface for the \textsc{Hipi} task.}
    \label{fig:hipi-question}
\end{figure}

\begin{figure}[t]
    \centering
    \includegraphics[width=\columnwidth]{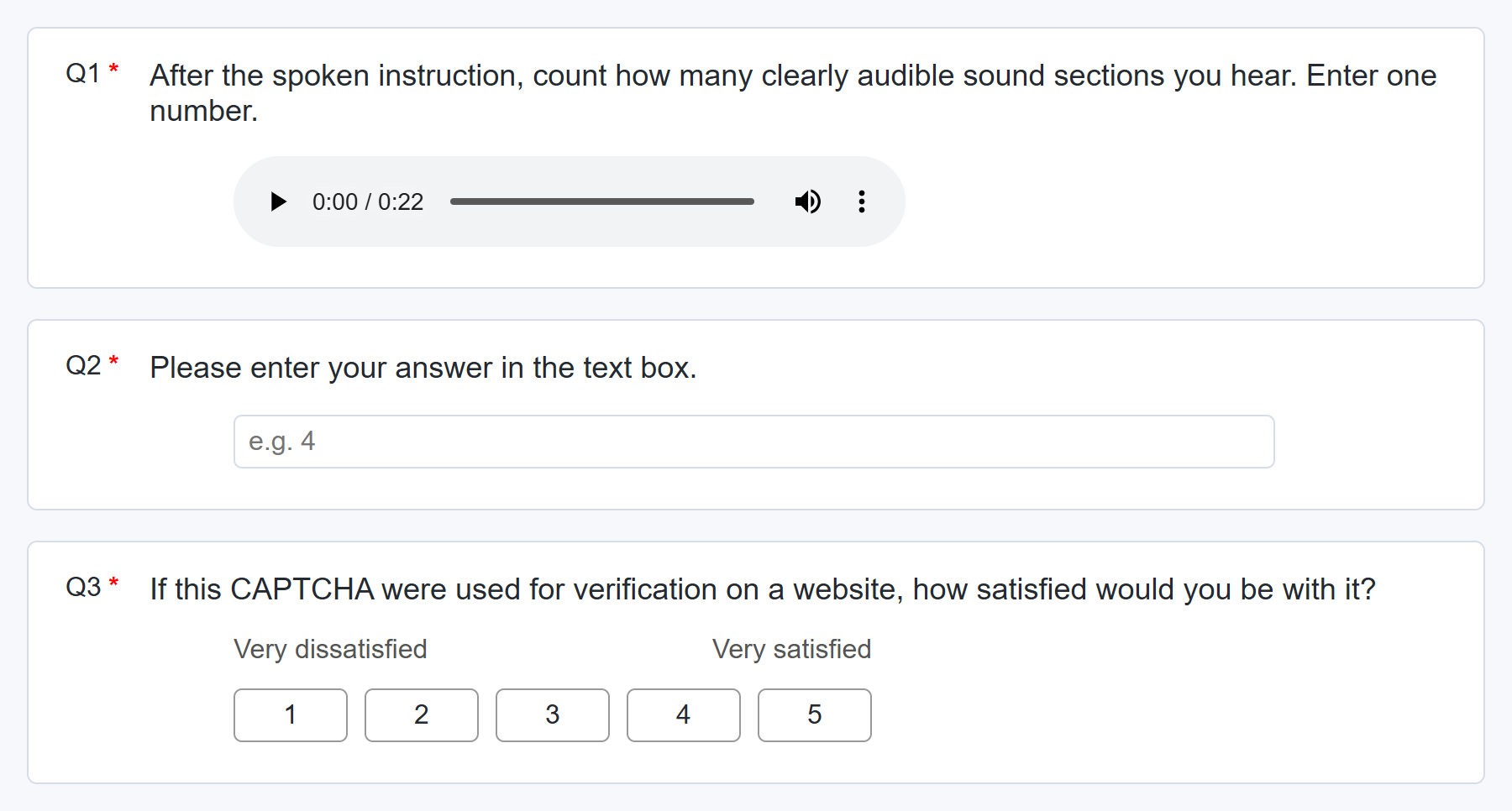}
    \caption{Participant interface for the \textsc{Yingchen} task.}
    \label{fig:yingchen-question}
\end{figure}
\section{Human Evaluation Materials}
\label{app:human_materials}

\subsection{Final-Evaluation Questionnaire}
\label{app:human-questionnaire}

{Participants were prohibited from using audio-analysis software, AI assistants, or other external tools while completing the challenges.}

{Figures~\ref{fig:hipi-question} and
\ref{fig:yingchen-question} show the participant-facing
interfaces for \textsc{Hipi} and \textsc{Yingchen},
respectively. Each interface contains three questions:
a task-specific instruction with an audio player,
a free-response answer field, and a satisfaction
rating asking how satisfied participants would be
if the CAPTCHA were used for website verification. 
}

\subsection{Demographic Information of Participants}
\label{app:distribution}

\begin{figure}[t]
    \centering
    \includegraphics[width=\columnwidth]{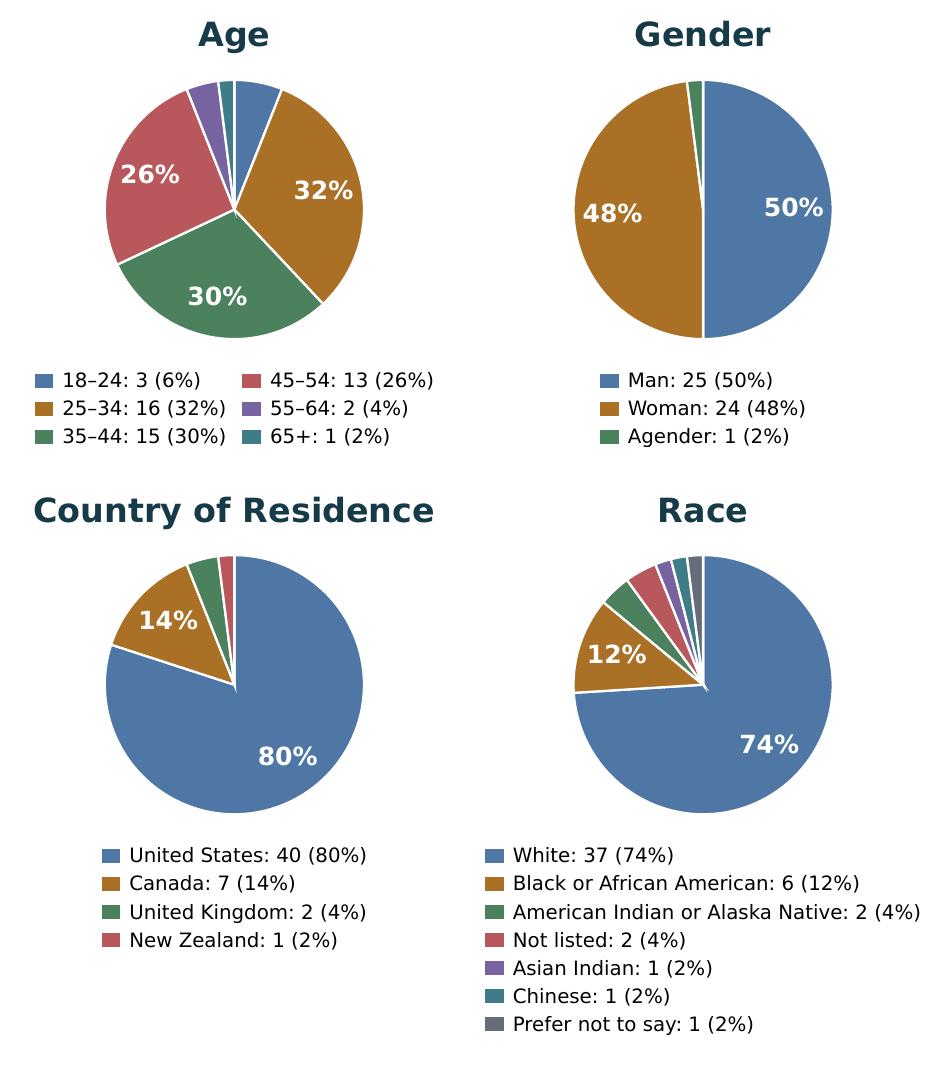}
    \caption{Aggregate demographic characteristics of the 50 participants in
    the final human evaluation.}
    \label{fig:demographics}
\end{figure}

Participants self-reported their age, gender, country of residence, and race.
We report these characteristics in aggregate to preserve participant privacy.
Figure~\ref{fig:demographics} summarizes the demographic characteristics of
all 50 participants who completed the final human evaluation. The age
distribution was 18--24 (6\%), 25--34 (32\%), 35--44 (30\%), 45--54 (26\%), 55--64 (4\%), and 65+ (2\%).
Participants identified as men (50\%), women (48\%), and agender (2\%).
The countries of residence were United States (80\%), Canada (14\%), United Kingdom (4\%), and New Zealand (2\%).
Responses to the race question included White (74\%), Black or African American (12\%), American Indian or Alaska Native (4\%), Asian Indian (2\%), Chinese (2\%), participants who selected an unlisted category (4\%), and participants who preferred not to report their race (2\%).

\subsection{Additional Human Acceptance Results}

Among the 50 participants, 38 passed both
\textsc{Hipi} and \textsc{Yingchen}, one passed only
\textsc{Hipi}, seven passed only \textsc{Yingchen},
and four passed neither. Of the 39 participants who
passed \textsc{Hipi}, 38 also passed \textsc{Yingchen},
yielding a composite HAR of 76\%. The 95\% Wilson score
confidence intervals~\cite{wilson1927} for HAR are
64.8--87.2\% for \textsc{Hipi}, 78.6--95.7\%
for \textsc{Yingchen}, and 62.6--85.7\% for their
composition. These results indicate substantial
human acceptance with limited additional acceptance
loss relative to \textsc{Hipi} alone.